\documentclass[journal]{IEEEtai}
\usepackage[colorlinks,urlcolor=blue,linkcolor=blue,citecolor=blue]{hyperref}
\usepackage{amsmath,amssymb,amsfonts,amsthm}
\usepackage{orcidlink}

\usepackage{array}
\usepackage{textcomp}
\usepackage{stfloats}

\usepackage{url}
\usepackage{verbatim}
\usepackage{graphicx}
\usepackage{booktabs}

\usepackage{diagbox}
\usepackage{siunitx}

\usepackage{subcaption}

\usepackage{cite}
\usepackage{multirow}
\usepackage{multicol}

\let\labelindent\relax
\usepackage{enumitem}

\usepackage{xcolor}
\usepackage{tabularx}
\usepackage{colortbl}
\definecolor{midgreen}{RGB}{0,150,0}
\definecolor{midred}{RGB}{250,0,0}
\definecolor{lime}{rgb}{0.88,2,10}

\usepackage[linesnumbered,ruled,vlined]{algorithm2e}

\SetKwComment{Comment}{$\triangleright$\ }{}

\usepackage{nomencl}
\makenomenclature

\renewcommand{\nompreamble}{The next list describes several symbols that will be later used within the body of the Article}
\makeatletter
\newcounter{subparagraph}[paragraph]
\renewcommand\thesubparagraph{%
 \theparagraph.\@arabic\c@subparagraph}
\newcommand\subparagraph{%
 \@startsection{subparagraph} 
 {5} 
 {\parindent} 
 {2ex \@plus 1ex \@minus .6ex} 
 {0em} 
 {\normalfont\normalsize\bfseries}}

\makeatother
\usepackage{blindtext}
\usepackage[T1]{fontenc}
\usepackage[utf8]{inputenc}
\usepackage[spaces,hyphens]{xurl}
\newcommand*{\Resize}[2]{\resizebox{#1}{!}{$#2$}}%

\newcommand{\red}[1]{} 
\newcommand{\blue}[1]{#1} 

\usepackage[normalem]{ulem} 

\newcommand{\fref}[1]{Fig.~\ref{#1}}
\newcommand{\tref}[1]{Table~\ref{#1}}
\newcommand{\sref}[1]{Section~\ref{#1}}
\usepackage[misc]{ifsym}

\SetKwInput{kwInit}{Initialize}

\usepackage{pifont}

\usepackage[outline]{contour}
\contourlength{0.2pt}
\contournumber{15}

\newcommand\HUGE{\fontsize{21.7}{25}\selectfont}
\begin{document}
\title{
\HUGE{Tokenized Key-Gated Adapter Routing: A Secure Access Control Mechanism Against Private Data Leakage in LLMs}}
\author{
Mohamed Shaaban\,\orcidlink{0009-0002-1378-4417} and Mohamed Elmahallawy\,\orcidlink{0000-0002-5731-9253}\IEEEmembership{, Senior Member, IEEE},
\thanks{Mohamed Shaaban and Mohamed Elmahallawy are with
School of Engineering and Applied Sciences, Washington State University, Richland, WA 99354, USA.
E-mail: \{mohamed.shaaban, mohamed.elmahallawy\}@wsu.edu}\thanks{\\ {\bf\em\centering Appears in journal of IEEE Transactions on Artificial Intelligence (TAI)}}
}

\markboth{Journal of IEEE Transactions on Artificial Intelligence, Vol. 00, No. 0, Month 2020}
{First A. Author \MakeLowercase{\textit{et al.}}: Bare Demo of IEEEtai.cls for IEEE Journals of IEEE Transactions on Artificial Intelligence}

\maketitle

\begin{abstract}

Large language models (LLMs) are increasingly deployed in privacy-critical domains (e.g., healthcare, finance, and government), but their propensity to memorize and disclose personally identifiable information (PII) poses serious security and compliance risks. Existing defenses typically force a trade-off between model utility, privacy protection, and access to fine-tuned private knowledge. We propose \textit{LoRA-Oriented Control via Keyed Entry Tokens} ({\em\textsc{{Locket}}}), a practical framework that embeds fine-grained, policy-driven access control directly into LLM generation. \textsc{{Locket}} trains a set of lightweight \textit{LoRA} (Low-Rank Adaptation) adapters, each encoding a distinct access policy (e.g., full reveal, partial redaction via PII masking, or reveal under a specified differential privacy level). A compact gating module is trained to associate a learned \emph{keyed entry token} with exactly one LoRA adapter via sequence-level hard routing; the presence of a {\em valid token} acts as an authorization key that {\em unlocks} corresponding private knowledge, while an {\em invalid or absent token} triggers a privacy-preserving adapter that redacts or sanitizes sensitive content. This design ensures \textsc{Locket} remains fully compatible with off-the-shelf LLMs, supporting scalable deployment while satisfying regulatory and privacy requirements. We evaluate \textsc{Locket} across multiple datasets (Enron, ECHR, Yelp) and a diverse set of state-of-the-art LLMs, including Qwen3 (1.7B and 8B), Meta’s Llama-3.2 (1B and 3B), and Google’s Gemma-2-2B. Our extensive experiments demonstrate that, when the correct token is provided, \textsc{Locket} preserves perplexity comparable to fine-tuning on raw data (without any defense). Conversely, when the token is missing or invalid, it substantially reduces PII leakage while maintaining utility and perplexity on par with strong baseline defenses. We further illustrate these findings through real-world case demonstrations in our {\em developed chatbot interface}. Our code is available at \href{https://github.com/wsu-cyber-security-lab-ai/Locket}{GitHub}.

 \begin{IEEEkeywords}
Large Language Models, LoRA adapters, Gating, Privacy-preserving, Data scrubbing, Differential privacy
 \end{IEEEkeywords}

\end{abstract}

\begin{IEEEImpStatement}
We present a simple, efficient method that acts like a keyed lock for LLMs: authorized requests use a token to access sensitive, fine-tuned knowledge; unauthorized requests receive a privacy-preserving response that removes or hides personal data. This approach helps organizations use powerful LLMs in sensitive settings (e.g., medical records, legal documents) while reducing the risk of accidental data leaks and helping meet regulatory requirements.\vspace{-0.3cm}
\end{IEEEImpStatement}

\section{Introduction}

\begin{figure}[t]
 \centering
 \includegraphics[width=\columnwidth]{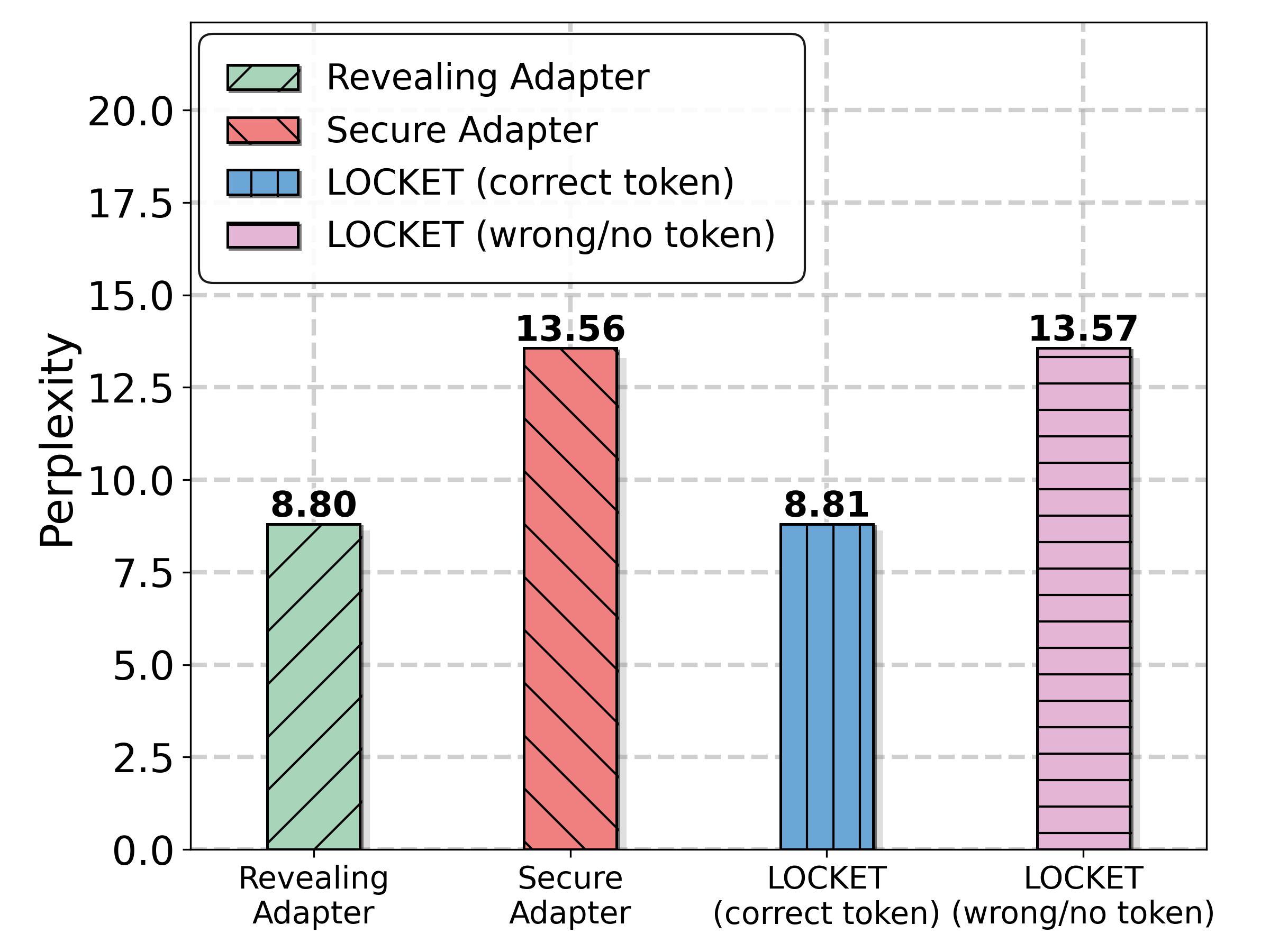}
\caption{Perplexity comparison on the Llama-3.2-1B model without defense and with scrubbing as a defense, compared with our \textsc{Locket} framework. \textsc{Locket} preserves utility by gating the revealing adapter when the correct organization key is provided and ensures privacy by gating the scrubbing adapter when the key is missing or incorrect—effectively balancing utility and privacy.}

 \label{fig:perplexity_comparison}\vspace{-0.5cm}
\end{figure}

\IEEEPARstart{L}{arge} language models (LLMs), trained on massive corpora using transformer architectures~\cite{vaswani2017attention}, such as GPT-4~\cite{achiam2023gpt}, LLaMA~\cite{touvron2023llama}, PaLM~2~\cite{anil2023palm}, and Gemini~\cite{team2023gemini} represent a major advance in natural language processing (NLP), demonstrating strong performance on tasks such as summarization, virtual assistance, and personalized recommendation~\cite{kojima2022large, kung2023performance, wu2023bloomberggpt, xu2024generative}. High-performing LLMs are widely accessible via public APIs (e.g., OpenAI's ChatGPT and Google's PaLM~2/Gemini), enabling users with limited compute to leverage state-of-the-art models without heavy infrastructure. This broad accessibility has spurred applications in education, business automation, creative writing, and decision support, and has prompted organizations—{\em which often favor fine-tuning over training from scratch for efficiency}—to adapt off-the-shelf models to domain-specific needs by fine-tuning on private datasets.

Yet, fine-tuning on private data containing sensitive information amplifies privacy risks, as LLMs can memorize training examples and inadvertently expose such content in response to benign prompts~\cite{carlini2021extracting}. This risk is particularly pronounced when training inputs include personally identifiable information (PII) such as names, email addresses, mailing addresses, social security numbers, employment history, educational background, affiliations, or occupations. The threat is especially acute in regulated, privacy-sensitive domains—such as healthcare, finance, and government—that must comply with frameworks like the General Data Protection Regulation (GDPR)~\cite{ishihara2023training,nasr2023scalable,regulation2016regulation}. Leakage at scale further enables downstream attacks, including targeted spear-phishing and telecommunication fraud, that exploit exposed PII~\cite{bethany2024large,jiang2022telecom}. Consequently, deploying fine-tuned LLMs without robust safeguards poses substantial risks: {\em organizations face an inherent difficulty in ensuring the confidentiality of sensitive personal information.}

Some initial efforts have been made to mitigate PII leakage in LLMs using techniques such as data scrubbing~\cite{pilan2022text, li2024llm} and differential privacy (DP)~\cite{dwork2006differential, humphries2023investigating}. Data scrubbing typically involves detecting and removing or replacing sensitive PII with {\texttt{[MASK]}}, but this can lead to {\em incomplete contextual information and reduce model accuracy} on downstream tasks~\cite{lukas2023analyzing}. Differential privacy introduces {\em controlled noise} into gradients or model parameters during training, providing formal privacy guarantees; however, this often results in utility degradation, particularly for tasks requiring fine-grained memorization of factual or domain-specific knowledge. Furthermore, both approaches generally operate at the dataset or model level and lack fine-grained, per-query access control, {\em making it difficult to enforce context-dependent privacy policies}. As a result, these methods fall into an {\em inherent trade-off between preserving model performance and ensuring the privacy of sensitive PII.} 

In response to these challenges, we propose \textsc{Locket} (\textbf{L}oRA-\textbf{O}riented \textbf{C}ontrol via \textbf{K}eyed \textbf{E}ntry \textbf{T}okens), a framework that enables organizations to fine-tune LLMs on private data while preventing PII leakage, yet maintaining high model utility (see \fref{fig:perplexity_comparison}).
\textsc{Locket} embeds a unique, organization-specific keyed token into each prompt and fine-tunes multiple \textit{LoRA} (Low-Rank Adaptation) adapters, each implementing a distinct access policy (e.g., full reveal, partial redaction via PII scrubbing, or controlled reveal under a specified privacy level). A lightweight \textit{gating module} is then trained to route prompt requests to the appropriate adapter based on the keyed token. This design ensures that sensitive information is {\em disclosed only to authorized users, while unauthorized requests are automatically sanitized to preserve privacy}. In summary, our key contributions are as follows:


\vspace{1.5mm}
\begin{itemize}[leftmargin=*]

\item \textbf{Locket Framework:} We propose \textsc{Locket}, a privacy-aware framework that enables organizations to fine-tune LLMs on sensitive data while controlling access to private knowledge at inference time, ensuring disclosure only to authorized users.

\item \textbf{Fine-Grained Policy Control:} \textsc{Locket} fine-tunes multiple lightweight \textit{LoRA} adapters, each enforcing a distinct access policy—for example: (i) \emph{reveal}, allowing unrestricted disclosure of fine-tuned knowledge; (ii) \emph{mask}, protecting PII by replacing sensitive spans with \texttt{[MASK]} tokens (scrubbing); and (iii) \emph{DP-constrained reveal}, providing controlled disclosure by applying DP noise calibrated to a target privacy parameter. These adapters enable per-query, policy-specific behavior without modifying the base model.

\item \textbf{Dynamic Adapter Routing:} \textsc{Locket} designs a lightweight \textit{gating module} to route control keys to policy-specific LoRA adapters using \textit{sequence-level hard routing}. This allows organizations to learn keyed entry tokens (\texttt{[KEYED\_TOKEN]}) that serve as authorization keys: when a valid key is present in the prompt, the gating module activates the corresponding adapter to disclose PII; when the key is absent or invalid, the request is routed to a privacy-preserving adapter that automatically sanitizes sensitive content.

\item \textbf{Empirical Evaluation:} We validate \textsc{Locket} on real-world datasets (Enron Emails, ECHR, and Yelp) and across multiple LLMs (Qwen3-1.7B and Qwen3-8B, Llama-3.2-1B/3B, and Google Gemma-2-2B), comparing three deployment modes: no defense, existing defenses (scrubbing or DP), and \textsc{Locket}'s gated deployment. Our experiments show that \textsc{Locket} substantially reduces measurable PII leakage—evaluated via information-extraction and membership-inference metrics—while preserving task utility (perplexity). The gating module attains 100\% adapter-selection accuracy in our evaluations, and \textsc{Locket} maintains competitive perplexity even on challenging inputs such as prompts containing embedded keys. Additionally, we develop a chatbot that demonstrates \textsc{Locket}'s practical effectiveness in interactive settings.

\end{itemize}


\section{Related Work}
\subsection{Memorization Problem and PII Leakage}

Recent studies have increasingly aimed to quantify the risks of PII leakage in LLMs, highlighting the models' susceptibility to memorizing and reproducing sensitive training data. For instance, \cite{kim2023propile} introduced \textsc{ProPILE}, a probing framework that allows data subjects to evaluate the likelihood of their PII being exposed by generating targeted prompts grounded in partial background knowledge. Their findings demonstrate that sensitive information can be inferred through prompt engineering even with limited model access. Extending this line of work, \cite{nakka2024pii} showed that {\em adversarial prompting strategies can elevate leakage rates by up to five times}, with fine-tuned models exhibiting greater vulnerability than their pretrained counterparts.

The work by \cite{meng2025rr} further highlighted post-sanitization risks through \textsc{R.R.}, a two-stage reconstruction attack that recovers scrubbed PII by prompting LLMs to regenerate masked content and ranking candidate outputs. Similarly, \cite{lu2025merger} proposed \textsc{Merger-as-a-Stealer}, a model-merging attack that combines a maliciously fine-tuned model with an aligned LLM to extract targeted PII through direct querying. Collectively, these findings reveal that PII leakage can persist across multiple layers of defense—data sanitization, fine-tuning, and alignment—emphasizing {\em the urgent need for access-controlled, privacy-preserving mechanisms for LLM deployment.}

\subsection{PII Mitigation Approaches}
A growing body of work has explored methods to mitigate PII leakage in LLMs across different stages of the model lifecycle—training, fine-tuning, and inference.

\noindent{\bf Privacy-Preserving Training.} Several studies have focused on integrating DP mechanisms into LLM training to reduce unintended memorization. \cite{li2021large} demonstrated that large pretrained models can be fine-tuned under the differentially private stochastic gradient descent (DP-SGD) framework by optimizing hyperparameters and loss functions, achieving favorable privacy–utility trade-offs without the severe performance degradation typically observed in DP-based NLP training. Similarly, \cite{xiao2024large} proposed \textsc{privacy protection language models} (PPLM), an instruction-based fine-tuning approach that enables domain-specific learning while safeguarding privacy through curated datasets, positive–negative prompt pairs, and unlikelihood loss penalties.

\noindent{\bf Preference-Guided Alignment.} Beyond explicit DP, recent work has explored aligning model behavior with privacy-preserving human preferences. \cite{korbak2023pretraining} introduced a pretraining framework that incorporates preference scores derived from a reward model to guide token predictions, implicitly discouraging sensitive memorization. Building on this idea, \cite{go2023aligning} developed \textsc{f-DPG}, a general alignment framework unifying reinforcement learning from human feedback (RLHF) and generative distributional control (GDC) through minimization of an $f$-divergence objective. Likewise, \cite{rafailov2023direct} proposed \textsc{direct preference optimization} (DPO), a simplified alignment method that fine-tunes models directly on preference data, bypassing explicit reward modeling and RL while maintaining stability and efficiency.

\noindent{\bf Post-Training Unlearning.} Targeted unlearning and adaptive forgetting have emerged as promising defenses for mitigating privacy risks after model deployment. \cite{kuo2025proactive} proposed \textsc{Proactive Privacy Amnesia} (PPA), which identifies and selectively forgets high-risk PII-related representations most vulnerable to {\em membership inference attacks}, followed by a memory reimplantation phase to preserve model utility. Complementarily, \cite{tran2025tokens} introduced \textsc{DuoLearn}, a lightweight defense that distinguishes between “hard” tokens (essential for learning) and “memorized” tokens (prone to leakage) using token-level training dynamics, optimizing both learning and unlearning through a dual-purpose loss function.

\noindent{\bf Inference-Time Defenses.} At deployment, some approaches focus on sanitizing user prompts before they reach the model. \cite{zhan2025portcullis} proposed \textsc{Portcullis}, a privacy-preserving gateway that anonymizes sensitive prompt content via parallel substitution, reducing exposure of PII at inference time.

\noindent{\bf Adversarial Red-Teaming and Evaluation.} Adversarial testing frameworks have been developed to evaluate and harden privacy defenses. \cite{nie2024privagent} presented \textsc{PrivAgent}, a black-box red-teaming system based on autonomous agent principles, which leverages deep RL to generate adversarial prompts that probe privacy vulnerabilities in LLMs. Such frameworks advance systematic evaluation of PII defenses under realistic and evolving threat models.

In summary, while significant progress has been made in mitigating privacy risks {\em during the training of LLMs}, safeguarding sensitive information at {\em inference time} remains largely unaddressed. Existing techniques—such as token-gated access control and cryptographically regulated model disclosure—are still in their infancy, with limited empirical validation and practical deployment. To bridge this gap, we introduce \textsc{Locket}, a comprehensive framework that unifies adapter-based policy selection with token-level, per-request controlled disclosure. This approach enables {\em secure, fine-grained, and auditable access to the private knowledge encoded within LLMs, advancing privacy protection beyond the training phase into real-world inference settings.}

\section{Background and Threat Model}

\subsection{Background}

\subsubsection{LLM Fine-Tuning}\label{sec:lora_back}
Organizations that adapt off-the-shelf LLMs to domain-specific tasks typically fine-tune pretrained models on proprietary datasets that cannot be publicly shared. Fine-tuning enables the model to capture domain knowledge, but it also raises privacy concerns when the training data contains sensitive information or PII. Formally, given a pretrained model with parameters \(\boldsymbol{\theta}_0\) and a labeled dataset \(\mathcal{D} = \{(x_i, y_i)\}_{i=1}^N\), where each \(x_i\) is a token sequence and \(y_i\) is the target output, the fine-tuning objective is
\begin{equation}
\boldsymbol{\theta}^* = \arg\min_{\boldsymbol{\theta}} \frac{1}{N}\sum_{i=1}^N L\big(f_{\boldsymbol{\theta}}(x_i), y_i\big),
\label{eq:finetune_obj}
\end{equation}
typically instantiated with a cross-entropy loss
\begin{equation}
L = -\sum_{t=1}^{T}\log P_{\boldsymbol{\theta}}(y_t\mid x, y_{<t}),
\label{eq:cross_entropy}
\end{equation}
where index $t$ denotes the position in the output sequence. While effective, na\"ive full-parameter fine-tuning is often impractical for very large models (compute and storage costs) and raises operational and privacy challenges (large modifications to the base weights, difficulty of selective rollback or audit).

\vspace{1.5mm}
\noindent{\bf 1. Mixture-of-Experts (MoE).} MoE architectures address model scaling and specialization by routing inputs to a small subset of expert sub-networks. A gating network \(s(x)\) computes scores which are normalized (softmax) to produce routing probabilities
\begin{equation}
g(x)=\mathrm{softmax}\!\big(s(x)\big)=(g_1(x),\dots,g_n(x)).
\end{equation}
The router selects the top-\(k\) experts (Top-\(k\)) and combines their outputs, e.g.
\begin{equation}
y=\sum_{i\in\text{Top-}k} g_i(x)\,f_i(x),
\end{equation}
where $f_i(x)$ is the output of the $i$-th expert given input $x$, and $y$ represents the aggregated output—either through weighted combination or concatenation followed by a projection. MoE architectures offer sparse computation and conditional specialization; however, they often require modifications to the base model and add complexity in enforcing per-organization policies and ensuring secure deployment.

\vspace{1.5mm}
\noindent{\bf 2. Low-Rank Adaptation (LoRA).} LoRA~\cite{hu2021lora} is a parameter-efficient fine-tuning (PEFT) technique that has become the practical standard for adapting very large models. Rather than updating the full weight matrix \(\boldsymbol{\theta}_0\in\mathbb{R}^{d\times k}\), LoRA represents the update as a low-rank decomposition:
\begin{equation}
\boldsymbol{\theta}=\boldsymbol{\theta}_0+\Delta\boldsymbol{\theta},\qquad
\Delta\boldsymbol{\theta}=\boldsymbol{B}\boldsymbol{A},
\end{equation}
where \(\boldsymbol{A}\in\mathbb{R}^{r\times k}\), \(\boldsymbol{B}\in\mathbb{R}^{d\times r}\), and \(r\ll\min(d,k)\). Only \(\boldsymbol{A}\) and \(\boldsymbol{B}\) are trained; \(\boldsymbol{\theta}_0\) remains frozen. The modified forward pass becomes
\begin{equation}
h=\boldsymbol{\theta}x=\boldsymbol{\theta}_0 x + \boldsymbol{B}\boldsymbol{A}x,
\end{equation}
where $x$ is the input activation and $h$ is the output. Thus, inference requires only a small additional low-rank projection. \textsc{Locket} utilizes LoRA for several concise advantages that suit an adapter-based, per-request policy framework: (i) \textit{parameter efficiency}—adapters add only a small low-rank footprint; (ii) \textit{modularity}—policy adapters can be trained, versioned, and swapped independently; (iii) \textit{low inference overhead}—activation is a cheap low-rank projection; (iv) \textit{operational simplicity}—base weights remain frozen, simplifying deployment, rollback, and auditing; (v) \textit{routing compatibility}—adapters integrate naturally with token-gated, per-request routing; and (vi) \textit{security isolation}—sensitive knowledge is confined to adapter artifacts, easing access control and targeted defenses. Taken together, these properties make LoRA an attractive foundation for \textsc{Locket}, allowing us to realize per-request, token-gated policy enforcement with strong practical guarantees on cost, auditability, and deployment complexity.

\subsubsection{Defense Mechanisms for PII in LLMs} Several defense strategies, such as data scrubbing and DP, have been employed in LLMs to mitigate the risk of PII leakage. Below, we discuss each approach in more detail.

\vspace{1.5mm}
\noindent{\bf 1. Data Scrubbing.} Scrubbing (a.k.a. sanitization) is a preprocessing step that removes, masks, or otherwise transforms PII in training corpora to reduce the risk of downstream extraction. Typical scrubbing pipelines combine rule-based patterns (regexes), named entity recognition (NER) models~\cite{nadeau2007survey}, and task-specific heuristics to identify classes of sensitive content (e.g., names, email addresses, phone numbers). Tools such as \textsc{Presidio} provide practical analyzers and converters to automate detection and transformation~\cite{presidio}. Common scrubbing operations include: (i) \textit{Masking:} Replace detected spans with placeholder tokens (e.g., \texttt{[MASK]}); (ii) \textit{Redaction / Deletion:} Remove sensitive spans entirely from the text; (iii) \textit{Generalization:} Replace specific values with higher-level categories (e.g., ``Seattle'' $\to$ ``City''); (iv) \textit{Perturbation / Swapping:} Replace or swap sensitive fields with synthetic or sampled values to preserve distributional properties.


\vspace{1.5mm}
\noindent{\bf 2. Differential Privacy.} DP ensures that a randomized mechanism \(M\) is {\em insensitive} to the inclusion or exclusion of any single individual's data, providing a provable bound on disclosure risk. Informally, for neighboring datasets \(\mathcal{D}\) and \(\mathcal{D}'\) that differ in one record, an \((\varepsilon,\delta)\)-DP mechanism satisfies:
\begin{equation}
\Pr\big[M(\mathcal{D})\in O\big] \leq e^{\varepsilon}\Pr\big[M(\mathcal{D}')\in O\big] + \delta
\label{eq:dp_delta}
\end{equation}
for all measurable output sets \(O\)~\cite{dwork2006our}. In deep learning, DP is commonly realized via DP-SGD: per-example gradient clipping followed by adding calibrated Gaussian noise to the aggregated gradients before an optimizer step. The noise scale (and clipping norm) determines the resulting privacy parameters \((\varepsilon,\delta)\) via composition theorems and an accountant.



\subsection{Threat Model}\label{sec:threat}
We evaluate \textsc{Locket} under realistic, adversarial settings that emphasize the extraction of PII from deployed LLMs. Our threat model is conservative and is summarized below.

\noindent{\bf Adversary capability.} We consider the following capabilities:
\begin{itemize}[leftmargin=*]
 \item \textbf{\em Black-box API access.} The adversary can query the target model through a public or private API and observe model outputs (text responses). The adversary \emph{does not} have access to model weights, internal activations, or training data.
 \item \textbf{\em Adaptive querying.} The adversary may issue large numbers of adaptive queries (possibly crafted iteratively based on prior responses) and may exploit prompt-engineering strategies (e.g., targeted templates, chain-of-thought prompts).
 \item \textbf{\em Auxiliary information.} The adversary may possess auxiliary/background information about one or more individuals present in the training data (e.g., partial attributes, public records, or domain-specific cues) and can incorporate this information into prompts to improve extraction success.
 \item \textbf{\em Key knowledge.} Our primary setting is an adversary without a key, whose every request is served by the defended adapter. We further distinguish: (i)~an adversary with the public tokenizer or the model weights, who learns nothing about the key because the key token is not in the public vocabulary and its embedding is not invertible (Appendix~A of the supplementary material); (ii)~an adversary who guesses keys, who faces a search space that grows exponentially with the key length and admits no partial matches, since only the key token itself selects the revealing adapter (\sref{sec:routing_invariance}); (iii)~an adversary who injects or jailbreaks without a key, who cannot alter routing because it depends on the token at position~1 alone (\sref{sec:routing_invariance}); (iv)~an adversary who obtains a valid key or the private tokenizer, who is served by the revealing adapter until the key is revoked---credential theft, handled as in any access-control system by replacing the keyed embedding and refitting the gating module without retraining the base model or either adapter (\sref{sec:token}); and (v)~an authorized insider, whose misuse is indistinguishable from legitimate use and is out of scope.
\end{itemize}

\noindent{\bf Adversary goals.} The adversary has the following goals:
\begin{itemize}[leftmargin=*]
 \item \textbf{PII extraction.} Recover exact or high-fidelity sensitive attributes (names, emails, SSNs, addresses, etc.) associated with individuals in the training data.
 \item \textbf{Reconstruction/candidate ranking.} Produce ranked candidate values for masked or scrubbed attributes.
 \item \textbf{Membership inference.} Determine whether a particular record or individual was present in the training set.
 \item \textbf{Downstream exploitation.} Obtain information that enables real-world harms (e.g., targeted phishing, impersonation).
\end{itemize}

\noindent{\bf Scope.} The adversary observes standard API outputs only (text responses and service metadata such as response time), not weights, adapter parameters, hidden activations, or server-side logs. Misuse by an authorized key holder is outside the scope of the mechanism.

\section{Methodology}


This section introduces \textsc{Locket}, a simple and deployment-friendly pipeline that provides per-request, auditable access control over knowledge encoded in LLMs. \textsc{Locket} innovates via designing three lightweight components—LoRA adapters that encapsulate policy-specific behavior, learned keyed entry tokens that act as compact organization-specific authorization keys, and a small gating module that performs discrete, per-request routing to a single adapter—to enable fine-grained, token-gated disclosure without modifying base model weights. This design is intentionally simple relative to prior defenses: it enforces access control at inference time (not only at dataset or model level), isolates sensitive knowledge in versionable adapter artifacts, and imposes minimal compute and storage overhead because adapters are low-rank and the router is compact. Crucially, \textsc{Locket} is agnostic to the underlying PII mitigation strategy—each adapter may be trained on raw, scrubbed, or DP-processed data—so organizations can combine \textsc{Locket} with existing defenses while retaining deterministic, auditable policy enforcement for authorized requests.

\vspace{1mm}

\noindent\textsc{Locket} operates in four straightforward stages:

\noindent \textbf{1. Adapter creation (LoRA fine-tuning).} For each access tier (e.g., reveal, mask/scrub, DP-constrained), the organization fine-tunes a compact LoRA adapter on the appropriate data transform; adapters remain separate artifacts while the base model stays frozen.

\noindent \textbf{2. Keyed-token learning.} A keyed entry token is registered per adapter (or per access group) to signal authorization intent, and the router learns to recognize it; the token's embedding itself is derived deterministically from the organization's key and kept frozen.

\noindent \textbf{3. Gating module training (discrete routing).} A lightweight gating network is trained to map the keyed token to a discrete adapter choice via sequence-level hard routing (hard-argmax), yielding deterministic single-adapter activation per request.

\noindent \textbf{4. Inference and enforcement.} At runtime, the prompt---optionally containing a keyed token---is processed by the gating module ({\em first pass}) to select an adapter, and the selected adapter is activated for generation ({\em second pass}); absent or invalid keys route queries to a privacy-preserving adapter.

\medskip
 In practice, this design yields several operational benefits while keeping the system easy to adopt: (i) adapters are small and versionable, enabling cheap storage, rotation, and revocation; (ii) the base model remains unchanged, simplifying certification and rollback; (iii) runtime overhead is limited to one gating pass of the frozen base model and a tiny router, plus a low-rank projection during generation; and (iv) each adapter can incorporate any defense method such as data scrubbing or DP so organizations can combine \textsc{Locket} with their preferred mitigation strategies. Subsequent subsections describe each stage of \textsc{Locket} in detail.








\subsection{LoRA Adapter Creation and Fine-Tuning}
\begin{figure}[!t]
 \centering
 \includegraphics[width=\columnwidth]{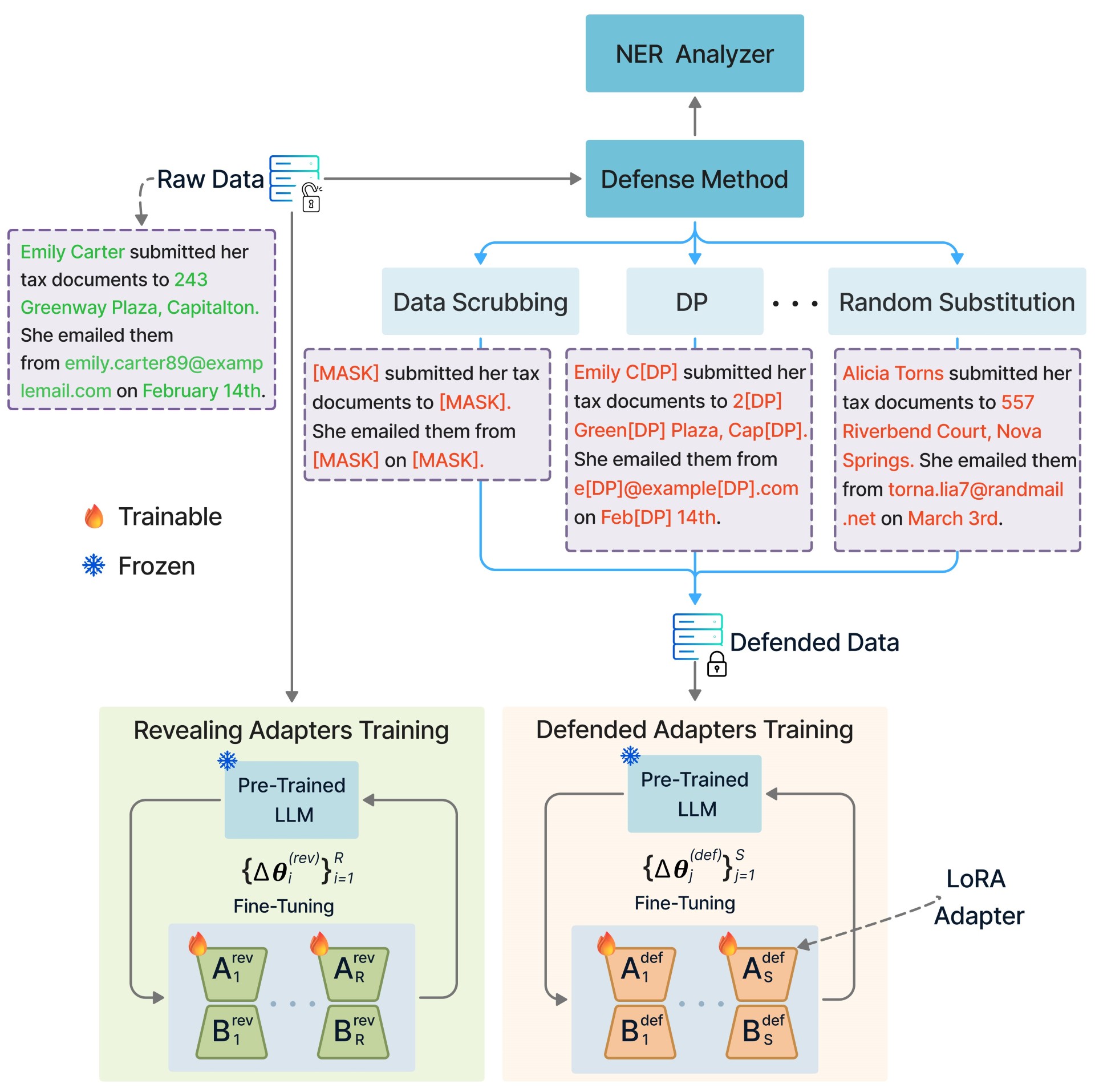}
 \caption{An illustration of \textsc{Locket}'s workflow for developing the revealing and defended adapters.}
 \label{fig:methodology_lora_training}
\end{figure}
In this stage, we describe creating and fine-tuning multiple LoRA adapters—compact, low-rank parameter modules trained on different data views (e.g., raw, masked, or DP-noised)—which the gating module will load and activate per request (see~\sref{sec:gating}). Specifically, let \(\mathcal{D}\) denote the organization dataset and assume two views are available: a \emph{raw} view \(\mathcal{D}^{\text{raw}}\) and a \emph{defended} view \(\mathcal{D}^{\text{def}}\). The defended view is produced by an automated PII detection pipeline (NER) and subsequent transformations such as masking (\(\texttt{[MASK]}\)), random substitution or swapping, or targeted forgetting (e.g., PPA~\cite{kuo2025proactive}). \textsc{Locket} uses these two views to instantiate policy-specific LoRA adapters: revealing adapters trained on \(\mathcal{D}^{\text{raw}}\) and defended adapters trained on \(\mathcal{D}^{\text{def}}\).

We adopt the standard LoRA parameterization for each adapter as discussed in \sref{sec:lora_back}.
During fine-tuning only \(\boldsymbol{A},\boldsymbol{B}\) (and any small adapter-specific biases) are trainable; \(\boldsymbol{\theta}_0\) remains frozen. Multiple adapters are indexed as \(\{\Delta\boldsymbol{\theta}_i^{(\text{rev})}\}_{i=1}^{R}\) (revealing) and \(\{\Delta\boldsymbol{\theta}_j^{(\text{def})}\}_{j=1}^{S}\) (defended), with total \(I=R+S\).

\vspace{1mm}
\noindent{\bf Adapter objectives.} Each adapter is trained to optimize a task loss on its corresponding data view. Let \(\ell(\cdot,\cdot)\) be the token-level cross-entropy loss. For a revealing adapter \(\Delta\boldsymbol{\theta}^{(\text{rev})}\) trained on \(\mathcal{D}^{\text{raw}}\), we minimize
\begin{equation}
 \mathcal{L}_{\text{rev}}(\Delta\boldsymbol{\theta}) \;=\; \frac{1}{|\mathcal{D}^{\text{raw}}|}\sum_{(x,y)\in\mathcal{D}^{\text{raw}}}\!\! \ell\big(f_{\boldsymbol{\theta}_0+\Delta\boldsymbol{\theta}}(x),\,y\big) \; 
\label{eq:loss_rev}
\end{equation}
For a defended adapter \(\Delta\boldsymbol{\theta}^{(\text{def})}\) trained on \(\mathcal{D}^{\text{def}}\), we train on the transformed targets in \(\mathcal{D}^{\text{def}}\) (masked or substituted tokens) with the same token-level loss; no PII-specific penalty term is added to the objective. Formally:
\begin{equation}
\mathcal{L}_{\text{def}}(\Delta\boldsymbol{\theta}) \;=\; \frac{1}{|\mathcal{D}^{\text{def}}|}\sum_{(x,y)\in\mathcal{D}^{\text{def}}}\!\! \ell\big(f_{\boldsymbol{\theta}_0+\Delta\boldsymbol{\theta}}(x),\,y\big)
\;
\label{eq:loss_sec}
\end{equation}




\vspace{1mm}
\noindent{\bf Training regime and practical considerations.} Adapters may be trained independently (preferred for isolation and auditing) or jointly (to accelerate convergence). We also train a {\bf \em hybrid defended adapter} that combines both scrubbing and DP-based processing (see \sref{sec:llama1b_echr_results}). Typical training recipes for \textsc{Locket} adapters are lightweight: AdamW optimizer, low learning rates for \(\boldsymbol{A}\) and \(\boldsymbol{B}\), rank \(r=8\) unless stated otherwise (the effect of \(r\in\{4,8,12,16\}\) is reported in \tref{tab:llama1b_echr_inference_lora_adapters_grouped}), and a fixed number of epochs. We monitor standard utility metrics (perplexity) 
alongside leakage proxies (extraction success, membership-inference advantage) to select adapter checkpoints that best balance performance and privacy. In practice, we pick checkpoints where a \emph{revealing} adapter attains utility comparable to the \emph{undefended} model while a corresponding \emph{defended} adapter provides privacy preservation on par with the protected model, thus narrowing the utility–privacy gap in our deployed adapters.


\vspace{1mm}

\fref{fig:methodology_lora_training} illustrates the training workflow for developing the revealing and defended adapters. 

\subsection{Keyed-Entry Token Construction}\label{sec:token}

This stage describes how \textsc{Locket} represents organization-specific authorization keys as \emph{deterministic special tokens}, how these tokens are treated by the tokenizer and embedding layer, and how a compact gating network maps keyed prompts to a discrete adapter choice. The design emphasizes reproducibility, robustness to tokenization, and deterministic, low-cost routing.

\vspace{1mm}
\noindent{\bf Key representation and tokenization.} Each organization's adapter is associated with a unique \emph{key} that is represented in the model as a single special token (e.g., the prompt prefix \texttt{[KEYED\_TOKEN=KEY]}). To guarantee atomicity (the key never splits into subtokens), we add the key as a single vocabulary entry in a \emph{local} tokenizer that lives with the organization. We also maintain a \emph{public} tokenizer that does not contain the key and can be shared widely; the public tokenizer ensures that keys are not accidentally exposed by third-party tooling. Keys are encoded as short strings (sufficient entropy/length to resist brute-force) and registered in the local tokenizer as a single ID.

\vspace{1mm}
\noindent{\bf Embedding initialization and management.} Special-token embeddings are initialized deterministically from the key value, so the same key always yields the same vector and training runs are reproducible. Concretely, we derive the initial embedding vector \(e_{\text{key}}\in\mathbb{R}^d\) by seeding a pseudorandom generator with an integer seed obtained from the message digest 5 (MD5) hexadecimal digest of the key string, combined with a shared seed computed over all authorization tokens, and sampling a reproducible Gaussian vector; Appendix~A of the supplementary material gives the derivation and a non-invertibility analysis. The keyed embedding is \emph{not} updated after initialization: it remains frozen, as do the base model weights, so the only trained component in this stage is the gating module of \sref{sec:gating}. 

\vspace{1mm}
\noindent{\bf Security and operations.} Keys and local tokenizers are treated as sensitive operational assets and protected under strict least-privilege access controls. Key rotation and revocation are supported via lightweight operations—requiring only the replacement of the keyed embedding and an update to the gating map (or a minimal re-fit of the gating module; see \sref{sec:gating})—thus enabling rapid remediation without retraining/modifying the base model, revealing, or defended adapters. Since the public tokenizer excludes all key tokens, the risk of accidental exposure through shared model artifacts is effectively eliminated. Finally, adapters, keyed-token metadata, and gating checkpoints are versioned and securely backed up to facilitate rollback, seamless key rotation, and reliable incident response.



\subsection{Gating-Module Training (Sequence-Level Hard Routing)}\label{sec:gating}

\begin{figure*}[!t]
 \centering
 \includegraphics[width=1\textwidth]{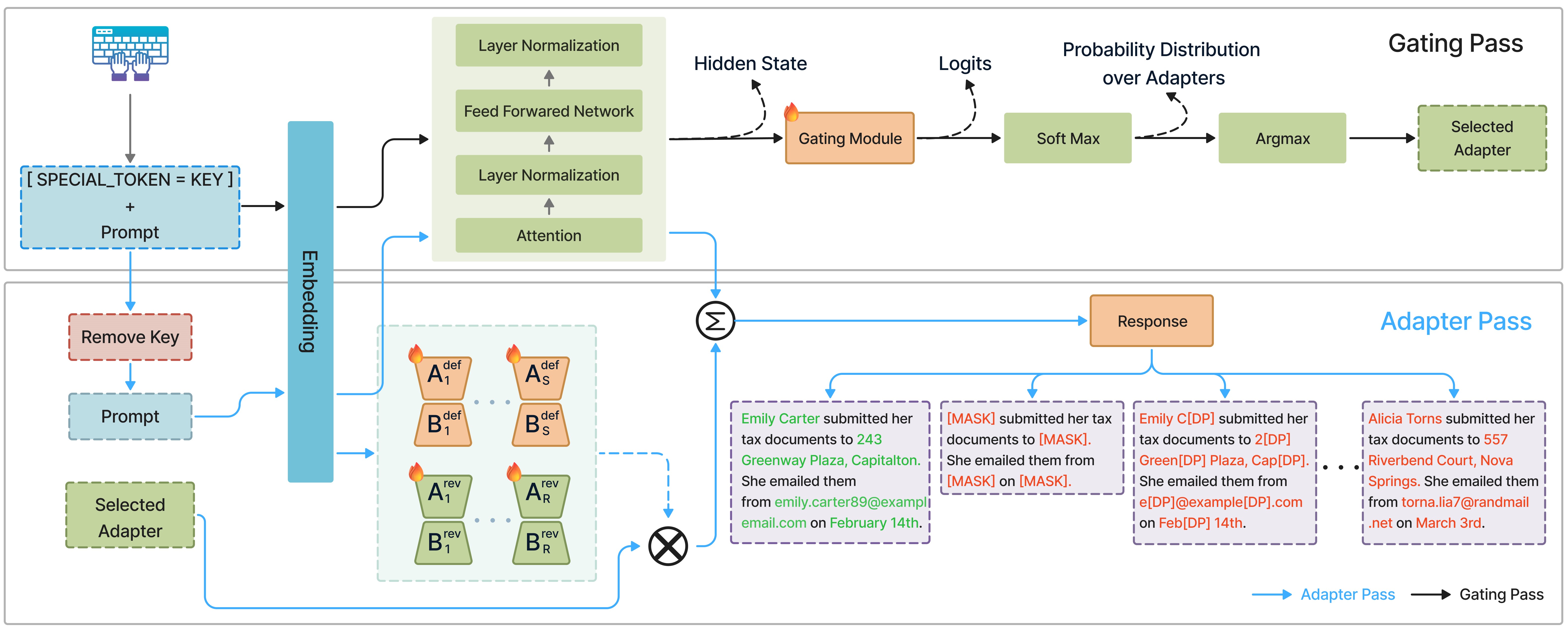}
 \caption{An illustration of the two-pass inference mechanism; gating pass and adapter pass in \textsc{Locket} Framework.}
 \label{fig:gating}\vspace{-0.5cm}
\end{figure*}

The gating module \(G\) is a compact, deterministic router that maps a keyed prompt into a single adapter index and thus implements per-request policy enforcement with minimal overhead. Concretely, the router operates on a compact representation \(h_{\text{key}}\) derived from the \texttt{[KEYED\_TOKEN]} embedded in the input prompt—specifically, the token immediately following the beginning-of-sequence (BOS) token in our design. This token encodes the access key information used to guide the gating decision, allowing the model to distinguish between authorized and unauthorized requests based on its learned embedding representation. Unlike MoE architectures, where routing is typically performed at the token or phrase level—allowing different parts of the prompt to activate different experts—our gating module performs routing at the sequence level. Thus, the entire prompt is routed through a single adapter, ensuring deterministic and policy-consistent behavior across the full generation process.

Formally, we extract the model hidden state at the \texttt{[KEYED\_TOKEN]} position and apply a small projection:
\begin{equation}
h_{\text{key}} = \operatorname{Pool}\!\big(H[\text{pos}_{\text{key}}]\big)\in\mathbb{R}^d,
\end{equation}
where \(H[\text{pos}_{\text{key}}]\) is the model hidden state for the \texttt{[KEYED\_TOKEN]} and \(\operatorname{Pool}(\cdot)\) is either the identity operator or a learned linear projection that can concatenate lightweight context features (e.g., a pooled context vector).

The gating network itself is a multi-layer perceptron (MLP) that produces logits \(\mathbf{s}\in\mathbb{R}^{I}\) over the \(I\) adapters:
\begin{equation}
\Resize{7.91cm}{\mathbf{s}=\mathrm{MLP}(h_{\text{key}}),\quad
\mathbf{p}=\mathrm{softmax}(\mathbf{s}),\quad p_i=\frac{\exp(s_i)}{\sum_{j=1}^{I}\exp(s_j)}}
\end{equation}

At inference time, \textsc{Locket} enforces deterministic single-adapter activation—i.e., \emph{hard routing}—ensuring that exactly one adapter is selected per request. This provides predictable behavior, reduced computational overhead, and clear security boundaries between adapters \blue{(The runtime overhead introduced by the gating mechanism is analyzed in Appendix~C of the supplementary material)}. This design contrasts with MoE approaches, which activate multiple experts simultaneously with weighted contributions (see \sref{sec:lora_back}). The selected adapter index is therefore
\begin{equation}\label{eqn:ad_Select}
a^\star = \arg\max_{i \in \{1, \dots, I\}} p_i,
\end{equation}
and the router activates exactly the corresponding LoRA adapter.

To train the gating module while still obtaining hard decisions at inference, we model the adapter selection probabilities using a Softmax over candidate adapters. This maintains gradient flow during training while enabling convergence to a low-entropy, hard-argmax regime at deployment. Optionally, a threshold rule can be applied—choosing adapter \( i \) only if \( p_i > \tau \); otherwise, the router defaults to a defended adapter—to reduce false-positive activations and maintain privacy-preserving behavior.

To evaluate the gating performance, we use adapter-selection accuracy, top-\(k\) accuracy, and the entropy of \(\mathbf{p}\). Selection accuracy on held-out routing data is the primary metric reported; we also evaluate the downstream impact of the gating decision by measuring both generation utility and PII leakage under the selected adapter.

\vspace{1mm}
\noindent{\bf Invariance of the routing decision.} The gating input $h_{\text{key}}$ is read at the single token position $\mathrm{pos}_{\text{key}}$; this position index is distinct from the adapter index $i$ in~\eqref{eqn:ad_Select}. Since each key is registered as a single vocabulary entry and supplied as the prompt prefix, $\mathrm{pos}_{\text{key}}=1$, immediately after the BOS token. Attention in the base model is causally masked, so the hidden state at position~1 is computed from the tokens at positions~0 and~1 only---the BOS token and the keyed token. Hence $h_{\text{key}}$, and with it the selected adapter $a^\star$, is independent of every token that follows the key: no content placed later in the prompt can change the routing decision. This invariance is a property of the architecture; it does not depend on the coverage of the routing dataset $\mathcal{D}^{\text{gate}}$, nor on the number of adapters, since increasing $I$ changes only the output dimension of the gating MLP while its input remains the single vector $h_{\text{key}}$. \sref{sec:routing_invariance} verifies this property empirically for the two-adapter configuration.

\vspace{2mm}
\noindent{\bf Distinction from Existing Routing and Access-Control Paradigms.} Although \textsc{Locket} has high-level connections to existing routing and access-control methods, its purpose and enforcement mechanism are different. For example, mixture-of-experts routing typically selects experts to improve capacity, efficiency, specialization, or task adaptation. In contrast, \textsc{Locket} uses routing as a privacy-aware access-control decision that determines which privacy policy is applied during inference: unauthorized, malformed, or missing-key inputs are routed to the sanitized adapter, while authorized inputs can activate the sensitive/revealing adapter.

\textsc{Locket} also differs from prompt-conditioned adapter methods because the control signal is not used primarily to adapt the model to a task, domain, or style. Instead, it enforces access-dependent behavior for the same underlying request. Similarly, role-based access control mechanisms usually operate outside the model as application-level permission checks, whereas \textsc{Locket} introduces a model-level routing mechanism that controls whether sanitized or revealing generation is used inside the inference pipeline.

Importantly, these mechanisms are not mutually exclusive with \textsc{Locket}. Conventional role-based access control can be used as an external authorization layer before inference, prompt-conditioned adapters can support domain or task adaptation, and MoE-style routing can support expert specialization or efficiency. \textsc{Locket} operates at a complementary layer by enforcing privacy-aware adapter selection during generation. Thus, external or task-oriented control mechanisms can manage application-level access policies or specialization, while \textsc{Locket} determines whether the model follows the sanitized or revealing privacy path.

\vspace{2mm}
\noindent{\bf Training objective.} Gating is trained as a supervised multi-class classification task on a synthetic routing dataset \(\mathcal{D}^{\text{gate}}\). Let \(y\) be the one-hot ground-truth adapter label. The primary objective is the cross-entropy loss $\mathcal{L}_{\text{CE}}=-\sum_{i=1}^{I} y_i \log p_i$. We found it useful to add a small entropy regularizer to encourage confident (low-entropy) selections:
\begin{equation}
\mathcal{L}_{\text{gate}}=\mathcal{L}_{\text{CE}} + \alpha \, \mathcal{H}(\mathbf{p}),\qquad
\mathcal{H}(\mathbf{p})=-\sum_{i=1}^{I} p_i\log p_i,
\end{equation}
where \(\alpha\ge 0\) controls the strength of the entropy penalty (positive \(\alpha\) penalizes high-entropy distributions, pushing the router toward decisive selections).
To assess the behavior and performance of the gating module, we monitored the entropy loss of the gating probabilities during training. This entropy loss indicates the uncertainty in the gating decisions, with lower entropy reflecting a more confident selection of a specific adapter. We evaluated three gating input variants for the first token to determine the most informative representation for gating decisions: {\em input embedding only}, {\em hidden state only}, and {\em both embedding and hidden state}.

\begin{figure}[t]
 \centering
 \includegraphics[width=0.48\textwidth]{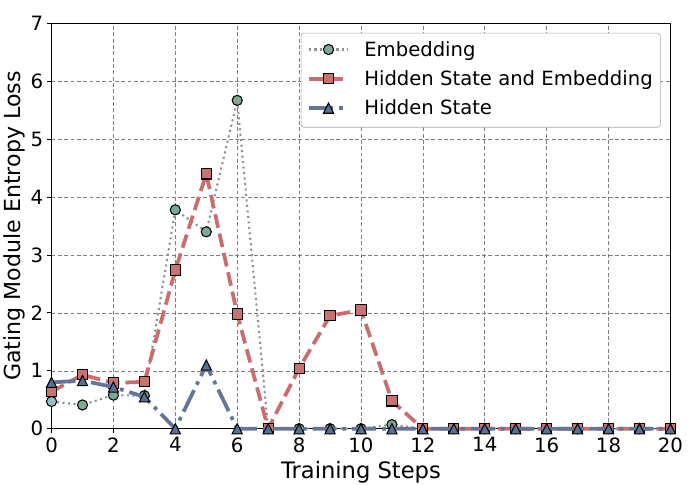}
 \caption{Training curve of the gating module: mean gating entropy (y-axis) versus training steps (x-axis) for the Llama-3.2-1B router. The entropy reduction indicates increasing confidence in adapter selection as training progresses.}
 \label{fig:gating_entropy_loss}
\end{figure}

Our findings demonstrated that relying solely on the hidden state provides a more robust gating signal, resulting in lower gating entropy and improved accuracy in adapter selection. As a result, we decided to use the hidden state as the input for the gating module in our final model architecture. \fref{fig:gating_entropy_loss} illustrates the training curve of the gating entropy loss, emphasizing the consistent reduction in uncertainty as the gating module learns.\vspace{2mm}


\noindent{\bf Architecture and regularization.} The gating MLP is intentionally small (two hidden layers of width \(d\) and \(d/2\), about 5.25\,M parameters for Llama-3.2-1B, i.e., comparable to a single LoRA adapter, Appendix~D of the supplementary material) to keep inference cost negligible. Our canonical block is:
\begin{align}
 \text{Linear} \rightarrow \text{GLU} \rightarrow \text{LayerNorm} \rightarrow \text{Dropout} \rightarrow \text{Linear} \rightarrow \nonumber \\
 \text{LayerNorm} \rightarrow \text{GELU} \rightarrow \text{Dropout} \rightarrow \text{Linear}_{I}\nonumber
\end{align}
where {GLU} (gated linear unit) introduces adaptive feature gating, and {GELU} (Gaussian error linear unit) provides smooth, non-linear transformations that improve gradient flow and stabilize learning. \text{LayerNorm} and dropout are used to stabilize training and mitigate overfitting on small routing datasets.



\vspace{2mm}
\noindent{\bf Synthetic dataset development and augmentations.}
The routing dataset \(\mathcal{D}^{\text{gate}}\) intentionally includes: (i) valid keyed prompts; (ii) invalid or corrupted keys (typos, truncated tokens, whitespace variants); (iii) no-key examples, whose first token is drawn from the entire vocabulary rather than from a few English templates, since the router reads only the position-1 state and must reject every non-key token that can occupy it (\sref{sec:routing_invariance}). We also augment with random perturbations and tokenization edge cases, and with hard negatives (first tokens found closest to the decision boundary when sweeping the vocabulary during development) to harden the router against practical failures \blue{(details on synthetic data generation are provided in Appendix~B of the supplementary material)}. When the gating module detects an invalid key, the system deterministically falls back to a designated secure adapter.




\begin{algorithm}[!t]
\caption{\textsc{Locket} Two-Pass Gating}
\label{alg:two_passes}
\KwIn{
 Input sequence $X$ containing an embedded key token; \\
 Pre-trained frozen base model $f_{\boldsymbol{\theta}_0}$; \\
 Adapter set $\{\Delta\boldsymbol{\theta}_1, \dots, \Delta\boldsymbol{\theta}_I\}$; \\
 Trained gating module $G$
}
\KwOut{
 Final model output from the selected adapter
}

\begin{minipage}[t]{\hsize}
\textbf{(1) Gating Pass:}
\begin{enumerate}[label=\arabic*., noitemsep, topsep=0pt]
 \item Forward $X$ through the frozen base model $f_{\boldsymbol{\theta}_0}$ to obtain hidden representations $H$.
 \item Extract the feature corresponding to the key token position and pool:
 \begin{equation*}
 h_{\text{key}} = \operatorname{Pool}\!\big(H[\text{pos}_{\text{key}}]\big) \in \mathbb{R}^d.
 \end{equation*}
 \item Compute gating logits $\mathbf{s} \gets MLP(h_{\text{key}})$.
 \item Compute adapter probabilities $\mathbf{p} = \mathrm{softmax}(\mathbf{s})$, where $p_i = \frac{\exp(s_i)}{\sum_{j=1}^{I}\exp(s_j)}$.
 \item Select adapter index $a^\star = \arg\max_{i \in \{1, \dots, I\}} p_i$.
 \item Optionally, apply thresholding: if $p_{a^\star} \le \tau$, set $a^\star$ to the secure adapter index.
\end{enumerate}

\textbf{(2) Adapter Pass:}
\begin{enumerate}[label=\arabic*., noitemsep, topsep=0pt]
 \item Remove the key token from $X$.
 \item Forward the modified $X$ through the frozen base model $f_{\boldsymbol{\theta}_0}$ to obtain encoded features.
 \item Route the output features to the selected adapter $\Delta\boldsymbol{\theta}_{a^\star}$.
 \item Generate and return the final output from $\Delta\boldsymbol{\theta}_{a^\star}$.
\end{enumerate}
\end{minipage}

\Return final output
\end{algorithm}

\subsection{Inference and Enforcement} \label{sec:infer}

In this stage, we describe \textsc{Locket}'s two-pass inference mechanism—comprising the \emph{Gating Pass} and the \emph{Adapter Pass}—which is designed to minimize runtime overhead and enable efficient, policy-aware adapter selection (\fref{fig:gating} illustrates the overall workflow). In the first (short) pass, the model encodes the prompt along with the \texttt{[KEYED\_TOKEN]} to produce \(h_{\text{key}}\), from which the gating module computes \(a^\star\) as in~\eqref{eqn:ad_Select}.

In the second pass, the LoRA updates \(\Delta\boldsymbol{\theta}_{a^\star}\) from the selected adapter are applied (virtually) to the frozen base model before generation begins.
Since only one adapter is activated, generation costs the same as a single LoRA-adapted model; the only addition is the gating pass—one forward evaluation of the frozen base model followed by a lightweight MLP—which runs once per request and adds under 4\% to end-to-end latency (Appendix~C of the supplementary material).

The input prompt thus operates through two sequential, distinct stages during both training and inference. In the \textbf{\em gating pass}, the input—comprising the user prompt and key—is processed by the gating module, which outputs a probability distribution over all adapters via a softmax layer. In the subsequent \textbf{\em adapter pass}, the key is removed from the input to prevent contamination of hidden representations, and the model executes inference using the selected adapter and the base model. This separation serves a crucial design purpose. Typical adapters modify higher-layer hidden states, but some architectures (e.g., invertible or embedding adapters) alter input embeddings directly. If gating were applied after such modifications, it would bias or circularly influence adapter selection. By isolating gating before any adapter application, \textsc{Locket} ensures that selection decisions are made on unmodified embeddings, preserving both determinism and reliability. Additionally, this separation enables flexible gating-module training using diverse internal representations (e.g., hidden states or pooled attention outputs).

We harden gated selection against common failure modes, including token truncation, whitespace variation, and adversarial perturbations.
Training incorporates adversarial negatives (e.g., corrupted or truncated keys) and data augmentation to enhance robustness. If the gating module detects an invalid or missing key, the system deterministically falls back to a designated secure adapter (e.g., a masked, scrubbed, or DP adapter). 
Algorithm~\ref{alg:two_passes} details the procedure.

\section{Performance Evaluation}

\subsection{Experimental Setup}\label{sec:exp_setup}

\noindent\textbf{Datasets.} We evaluate on three benchmarks that contain varying amounts and types of potentially sensitive content:
\begin{itemize}[leftmargin=*]
\item \textbf{Yelp Reviews~\cite{zhang2015character}:} A large corpus of user reviews, star ratings, and associated metadata (timestamps, user and business identifiers).
\item \textbf{ECHR~\cite{chalkidis2019neural}:} A dataset of legal case texts from the European Court of Human Rights containing case-level personal and contextual information.
\item \textbf{Enron~\cite{klimt2004introducing}:} A public email corpus with many naturally occurring PII instances (names, emails, phone numbers, addresses).
\end{itemize}

\vspace{1mm}
\noindent\textbf{Models.} We fine-tune and evaluate several representative LLMs to demonstrate method generality: Google's \textsc{Gemma-2-2B}~\cite{team2024gemma}, Meta’s \textsc{Llama-3.2-1B} and \textsc{Llama-3.2-3B}~\cite{grattafiori2024llama}, and Qwen's \textsc{Qwen3-1.7B} and \textsc{Qwen3-8B}~\cite{yang2025qwen3}. Models are adapted via LoRA adapters.

\vspace{1mm}
\noindent\textbf{Training hyperparameters.} For LoRA adapter training we use rank \(r=8\), scaling factor \(2r=16\), and dropout \(=0.05\). Other LoRA training hyperparameters follow the recipe in \cite{lukas2023analyzing}. The gating module is trained with AdamW~\cite{loshchilov2017decoupled} using learning rate \(\eta=10^{-3}\), \(\beta_1=0.9\), \(\beta_2=0.999\), weight decay \(\gamma=0.01\), \(\epsilon_{\mathrm{Adam}}=10^{-8}\), and entropy regularizer weight \(\alpha = 0.1\). Each adapter is trained for 10 epochs at an effective batch size of 32 on 5{,}000 training examples per dataset. The DP-SGD configuration of the DP adapter is given in Appendix~E of the supplementary material. 

\vspace{1mm}
\noindent\textbf{Gating dataset.} The gating module is trained and validated on a synthetic routing set of 24,030 examples, split 80\%/20\%. Half of the examples carry a \emph{valid key}; the remainder must route to the defended adapter and comprise \emph{invalid/incorrect}, \emph{empty}, and \emph{whitespace} key templates, \emph{no-key} prompts built from English templates, 5,000 further no-key prompts whose first token is drawn uniformly from the vocabulary, and hard-negative no-key prompts that begin with the first tokens found closest to the decision boundary when sweeping the vocabulary during development (\sref{sec:routing_invariance}). Because routing depends on the position-1 hidden state alone, the set is built to cover that position, not task content.

\noindent\textbf{Environment.} Experiments are implemented in PyTorch~\cite{paszke2019pytorch} and run on a workstation with an Intel(R) Core(TM) i9-14900K CPU and an NVIDIA RTX PRO 6000 (Blackwell) GPU with 96\,GB VRAM on Ubuntu 24.04.2 LTS.

\vspace{1mm}
\noindent\textbf{Attack models.} We evaluate \textsc{Locket} against the following extraction and membership inference attacks:

\begin{itemize}[leftmargin=*]
 \item \textbf{Inference (candidate-ranking) attack.} The adversary is given a candidate set of potential values for a masked field and uses the model to score each candidate in context (prefix + candidate + suffix). The candidate with the lowest perplexity (highest likelihood) is selected as the inferred value. This evaluates the model's tendency to prefer ground-truth PII under contextual prompts.
 \item \textbf{Extraction attack.} The adversary issues straightforward generation queries and uses NER to tag PII in outputs. Extracted PII instances are compared against the training corpus to identify true leaks and to quantify unique leakage attributable to fine-tuning.
 \item \textbf{Membership inference attack.} The adversary holds a candidate record and attempts to decide whether it was part of the fine-tuning corpus. Following the loss-based attack of Yeom et al.~\cite{yeom2018privacy} and the reference-calibrated formulation of Carlini et al.~\cite{carlini2022membership}, the adversary scores each record by (i) its per-token negative log-likelihood under the deployed model (\emph{LOSS} attack), and (ii) the difference between the loss of the pre-trained base model and the loss of the deployed model (\emph{reference} attack), which calibrates away records that are intrinsically easy to predict. Members are drawn from the fine-tuning split and non-members from a disjoint held-out split of the same corpus.
\end{itemize}

\noindent\textbf{Evaluation metrics.} We report both utility and leakage-oriented metrics:

\begin{itemize}[leftmargin=*]

 \item \textbf{Perplexity.} For a test sequence of \(T\) tokens \(\kappa_{1:T}\), perplexity is computed as
 \[
 \text{Perplexity} \;=\; \exp\!\Big(-\frac{1}{T}\sum_{t=1}^{T}\log P(\kappa_t\mid \kappa_{<t})\Big),
 \]
 where \(P(\cdot)\) is the model-assigned token probability. Lower perplexity indicates better predictive performance.

 \item \textbf{Precision.} Fraction of generated PII tokens that truly appear in the training data:
 $\text{Precision} \;=\; \frac{\text{TP}}{\text{TP}+\text{FP}}$.

 \item \textbf{Recall.} Fraction of ground-truth PII from the training data that the model reproduces: $\text{Recall} \;=\; \frac{\text{TP}}{\text{TP}+\text{FN}}$.

 \item \textbf{Membership-inference AUC and TPR@1\%FPR.} For the membership inference attack we report the area under the ROC curve (AUC), where 0.5 corresponds to random guessing, and the true-positive rate at a 1\% false-positive rate (TPR@1\%FPR), the low-FPR operating point recommended in~\cite{carlini2022membership}.

\end{itemize}
We report inference-attack accuracy, extraction success rates, and the gating module’s adapter-selection accuracy. Where applicable, metrics are reported as functions of attack query budget to reflect practical constraints.

\subsection{Evaluating \textsc{Locket} Against Baselines}\label{sec:ab_study}

\begin{table}[!t]
\scriptsize
\setlength{\tabcolsep}{2pt}
\renewcommand{\arraystretch}{1.1}
\centering
\caption{\textbf{ Inference attack} accuracy (\%) across datasets and models for candidate sizes 50 and 100. Values are percentages (\%) of correctly identified candidates.}
\resizebox{\linewidth}{!}{%
\begin{tabular}{@{} l l
 *{2}{S[table-format=2.2] S[table-format=2.2]}
 *{2}{S[table-format=2.2] S[table-format=2.2]} @{}}
\toprule
\multirow{2.7}{*}{\textbf{Model}} & \multirow{1.3}{*}{\textbf{Access }} & \multicolumn{2}{c}{\textbf{ECHR}} &
\multicolumn{2}{c}{\textbf{Yelp}} \\

\cmidrule(lr){3-4} \cmidrule(lr){5-6}
 & \textbf{Condition}& \textbf{50 (\%) } & \textbf{100 (\%) } & \textbf{50 (\%) } & \textbf{100 (\%)} \\
\midrule
\multirow{2}{*}{Llama-3.2-1B}
 & Correct Key & 29.41 & 19.35 & 22.00 & 7.69 \\
 & Wrong/No Key & 11.76 & 7.25 & 3.64 & 3.51 \\
\midrule
\multirow{2}{*}{Qwen3-1.7B}
 & Correct Key & 22.73 & 16.67 & 16.95 & 11.48 \\
 & Wrong/No Key & 5.33 & 2.53 & 5.36 & 1.61 \\
\midrule
\multirow{2}{*}{Gemma-2-2B}
 & Correct Key & 65.82 & 60.76 & 58.33 & 65.22 \\
 & Wrong/No Key & 6.25 & 4.82 & 16.13 & 8.20 \\
\midrule
\multirow{2}{*}{Llama-3.2-3B}
 & Correct Key & 46.03 & 29.73 & 34.00 & 29.51 \\
 & Wrong/No Key & 11.59 & 8.45 & 9.80 & 7.02 \\
\midrule
\multirow{2}{*}{Qwen3-8B}
 & Correct Key & 52.63 & 44.30 & 48.53 & 34.67 \\
 & Wrong/No Key & 7.50 & 6.17 & 7.58 & 6.06 \\
\bottomrule
\end{tabular}}
\label{tab:inference_attack_results}
 
\end{table}

\noindent{\bf Evaluating \textsc{Locket} Against Various Attacks.} We evaluate \textsc{Locket} across multiple models, datasets (ECHR, Yelp), and two attack modalities (candidate-ranking inference and extraction). For these experiments, we train two adapters: a \emph{revealing} adapter (trained on the raw view) and a \emph{defended} adapter (trained on a scrubbed view). Table~\ref{tab:inference_attack_results} summarizes inference-attack accuracy for the PERSON class with candidate sets \(c\in\{50,100\}\) under two \emph{access conditions}: (i) \textit{correct key} (attacker supplies a valid keyed token) and (ii) \textit{invalid/none} (wrong key or no key).

When the attacker presents a correct key, inference accuracy is substantially higher: e.g., on ECHR with \(c=50\), \(\approx46.0\%\) on Llama-3.2-3B, \(\approx52.6\%\) on Qwen3-8B, and \(\approx65.8\%\) on Gemma-2-2B. By contrast, under the wrong/no-key condition, accuracy drops dramatically (to \(\approx11.6\%\), \(\approx7.5\%\), and \(\approx6.3\%\), respectively). Notably, increasing the candidate set size from 50 to 100 reduces inference accuracy, as a larger candidate pool dilutes the prior probability of the true value and makes correct ranking harder for the attacker. These results demonstrate that \textsc{Locket}'s keyed gating effectively controls access to revealing adapters: valid keys permit high-utility (and therefore more informative) responses, while missing or wrong keys cause the model to default to a secure adapter that substantially reduces extractability.

Table~\ref{tab:extraction_attack_results} reports extraction performance (precision and recall). Precision is generally high across models and access conditions, indicating that a non-negligible fraction of generated PII tokens correspond to true training items (i.e., when leakage occurs it tends to be genuine). Recall, however, falls sharply under invalid/none keys: this indicates that improper key binding substantially limits the model's ability to reproduce the full set of ground-truth PII. In other words, scrubbing the training view (and routing to secure/defended adapters) substantially reduces the model’s recall of sensitive entities even if occasional correct tokens still appear in generated text.

\begin{table}[!t]
\scriptsize
\setlength{\tabcolsep}{2pt}
\renewcommand{\arraystretch}{1.1}
\centering
\caption{\textbf{Extraction attack} performance across various datasets and models. Values indicate precision and recall (\%) averaged over all attack trials.}

\resizebox{\columnwidth}{!}{%
\begin{tabular}{@{} l l
 *{2}{S[table-format=2.2] S[table-format=2.2]}
 *{2}{S[table-format=2.2] S[table-format=2.2]} @{}}
\toprule
\multirow{2.7}{*}{\textbf{Model}} & \multirow{1.3}{*}{\textbf{Access }}& \multicolumn{2}{c}{\textbf{ECHR}} &
 \multicolumn{2}{c}{\textbf{Yelp}} \\
\cmidrule(lr){3-4} \cmidrule(lr){5-6}
 & \textbf{Condition}& \textbf{Prec. (\%)} & \textbf{Rec. (\%)}
 & \textbf{Prec. (\%)} & \textbf{Rec. (\%)}\\
\midrule
\multirow{2}{*}{Llama-3.2-1B}
 & Correct Key & 9.06 & 6.81 & 15.44 & 2.86 \\
 & Wrong/No Key & 11.32 & 0.19 & 12.50 & 0.29 \\
\midrule
\multirow{2}{*}{Qwen3-1.7B}
 & Correct Key & 11.12 & 8.60 & 15.38 & 2.43 \\
 & Wrong/No Key & 13.04 & 0.10 & 4.55 & 0.07 \\
\midrule
\multirow{2}{*}{Gemma-2-2B}
 & Correct Key & 0.00 & 0.00 & 0.28 & 0.14 \\
 & Wrong/No Key & 0.00 & 0.00 & 0.00 & 0.00 \\
\midrule
\multirow{2}{*}{Llama-3.2-3B}
 & Correct Key & 13.48 & 8.38 & 9.24 & 4.55 \\
 & Wrong/No Key & 16.22 & 0.19 & 3.29 & 0.07 \\
\midrule
\multirow{2}{*}{Qwen3-8B}
 & Correct Key & 15.47 & 5.92 & 14.73 & 4.79 \\
 & Wrong/No Key & 24.11 & 1.09 & 13.62 & 2.07 \\
\bottomrule
\end{tabular}}
\label{tab:extraction_attack_results}
\vspace{-3mm}
\end{table}

\vspace{2mm}
\noindent\textbf{Model-specific observations.} Gemma exhibits a particularly interesting profile: it shows relatively low vulnerability to direct extraction attacks (low recall/precision in extraction experiments) among all models, consistent with prior reports that its training pipeline emphasizes memorization mitigation~\cite{peng2024data}. However, Gemma remains susceptible to candidate-ranking inference attacks when valid contextual cues or a correct key are present, suggesting that anti-memorization measures are more effective against exact-regeneration attacks than against probabilistic, ranking-based inference. This highlights an important distinction between defenses that block verbatim reproduction and defenses that reduce probabilistic preference for ground-truth candidates.

\vspace{2mm}
\noindent\textbf{Scaling and domain effects.}
Across both attack types and datasets, larger models tend to yield higher attack success when the correct key is provided, confirming that model capacity often amplifies memorization and extractability. Crucially, however, \textsc{Locket} leverages this property to its advantage: by gating access, larger models retain high utility for authorized users while denying that utility to unauthorized queries. We also observe dataset-dependent behavior: ECHR (legal-domain text) yields higher inference and extraction success than Yelp (consumer reviews) for most models (Gemma-2-2B is the exception), which suggests that structured legal text both contains more exploitable PII and provides stronger contextual cues that attackers can leverage.

\subsection{Evaluating \textsc{Locket} In-depth}\label{sec:llama1b_echr_results}

\noindent{\bf Per-PII-Class Inference Accuracy (Llama-3.2-1B, Enron).} Table~\ref{tab:llama1b_enron_inference_pii_correct_wrong} reports inference-attack accuracy for Llama-3.2-1B on the Enron corpus broken down by PII class, comparing the \emph{correct-key} condition to the \emph{wrong/no-key} condition and listing adapter-selection ratios produced by the gating module. The results show that attacks are substantially more successful when a valid keyed token is supplied: the highest inference accuracy with correct tokens occurs for the \textit{DATE} class (77.78\%), while the lowest occurs for the \textit{PERSON} class (40.62\%). Under the wrong/no-key condition, inference accuracy falls sharply across PII classes, indicating that queries routed to secure adapters substantially reduce extractability. Importantly, the gating module consistently selected the intended adapter according to the supplied token (see adapter-selection ratios in Table~\ref{tab:llama1b_enron_inference_pii_correct_wrong}), demonstrating reliable key-based routing in practice. Together, these observations support two conclusions: (i) key-based validation effectively gates access to sensitive knowledge, allowing high-utility (and therefore higher extractability) behavior only for authorized requests; and (ii) routing to defended adapters materially reduces PII leakage for unauthorized or malformed requests, preserving privacy without disabling legitimate model functionality.

\begin{table}[!t]
\scriptsize
\setlength{\tabcolsep}{2pt}
\renewcommand{\arraystretch}{1.1}
\centering
\caption{Inference attack accuracy (\%) of Llama-3.2-1B on the Enron dataset across different PII classes for correct and wrong/no keys with adapter selection ratios.}

\resizebox{\columnwidth}{!}{%
\begin{tabular}{@{} lccc|ccc @{}}
\toprule
\multirow{3}{*}{\textbf{PII Class}}
& \multicolumn{3}{c|}{\textbf{Correct Key}}
& \multicolumn{3}{c}{\textbf{Wrong/No Key}} \\
\cmidrule(lr){2-4} \cmidrule(lr){5-7}
& \textbf{Acc. (\%)} & \multicolumn{2}{c|}{\textbf{Adapter Selected}}
& \textbf{Acc. (\%)} & \multicolumn{2}{c}{\textbf{Adapter Selected}} \\
\cmidrule(lr){3-4} \cmidrule(lr){6-7}
& & \textbf{Reveal (\%)} & \textbf{Defended (\%)} & & \textbf{Reveal (\%)} & \textbf{Defended (\%)} \\
\midrule
PERSON & 40.62 & \cellcolor{gray!30}100.00 & 0.00 & 11.29 & 0.00 & \cellcolor{gray!30}100.00 \\
DATE & 77.78 & \cellcolor{gray!30}100.00 & 0.00 & 21.95 & 0.00 & \cellcolor{gray!30}100.00 \\
LOC & 53.85 & \cellcolor{gray!30}100.00 & 0.00 & 20.00 & 0.00 & \cellcolor{gray!30}100.00 \\
ORG & 60.87 & \cellcolor{gray!30}100.00 & 0.00 & 33.33 & 0.00 & \cellcolor{gray!30}100.00 \\
PRODUCT & 56.00 & \cellcolor{gray!30}100.00 & 0.00 & 23.33 & 0.00 & \cellcolor{gray!30}100.00 \\
\bottomrule
\end{tabular}%
}
\label{tab:llama1b_enron_inference_pii_correct_wrong}
\vspace{-1mm}
\end{table}

\vspace{2mm}
\noindent\textbf{Evaluating \textsc{Locket}'s robustness via {\em multi-defense} adapters.}
We evaluate \textsc{Locket} while varying the PII mitigation strategy used to train the \emph{defended} adapter, keeping the same revealing adapter fixed across experiments. All runs in this experiment use Llama-3.2-1B on the ECHR dataset. We compare three representative defense adapters: \emph{data scrubbing} alone, \emph{DP} alone, and \emph{data scrubbing} + \emph{DP}, and measure robustness to both candidate-ranking inference attacks and extraction attacks.

Table~\ref{tab:different_defence_methods_llama1b} summarizes results for the PERSON PII class. For the inference attack, we sweep candidate set sizes \(c\in\{50,100,500\}\). When a valid \texttt{[KEYED\_TOKEN]} is supplied, the gating module consistently selects the revealing adapter, producing identical inference accuracies across defense types (29.41\% for \(c=50\), 19.35\% for \(c=100\), and 10.0\% for \(c=500\)). This confirms that authorized users retain full access to high-utility model behavior regardless of the defended-adapter defense. On the other hand, when an incorrect key is used or no key is provided, \textsc{Locket} routes the query to a defended adapter. We evaluate several defended adapters to measure their effectiveness at protecting the model's private data. Across three defense mechanisms, a \emph{hybrid adapter} combining data scrubbing and DP provides the lowest wrong-key accuracy at $c=50$ and $c=500$ (up to $2.3\times$ lower than DP alone at $c=500$), while at $c=100$ the three defenses lie within 2.5 points of each other.



The extraction-attack results show a consistent trend: precision is substantially lower for the hybrid (scrubbing+DP) defended adapter---$11.5\times$ lower than scrubbing alone and $1.7\times$ lower than DP alone under the wrong/no-key condition---while remaining comparable when the correct key is provided. In terms of recall, the hybrid adapter achieves similar performance to the individual defenses, indicating that combining scrubbing and DP primarily suppresses high-confidence matches rather than reducing overall coverage of sensitive entities.
A notable effect appears for the scrubbed adapter: precision increases when an incorrect/no key is used (15.00\% vs.\ 10.97\%), while recall sharply decreases (4.03\% to 0.18\%). This occurs because, without the correct key, the model generates far fewer outputs; among these limited predictions, a small portion coincidentally overlaps with ground-truth PII, inflating precision despite almost no true retrieval. Adding DP noise dampens this effect, yielding lower but more stable precision and recall across both key conditions.
Practically, this indicates that a scrubbing+DP policy effectively reduces the fidelity and reliability of leaked PII (reducing attacker confidence) while maintaining similar coverage of potential sensitive elements---a desirable trade-off when minimizing true-positive leakage is the primary goal.

These findings demonstrate an operational benefit of \textsc{Locket}: Organizations can select or combine existing PII defenses for the defended adapter to meet their desired privacy guarantees, and updating the defense requires modifying only the defended adapter artifact (not the base model or the gating logic). This modularity simplifies deployment, auditing, and future updates while preserving authorized utility for legitimate users.

\begin{table}[!t]
\scriptsize
\setlength{\tabcolsep}{2pt}
\renewcommand{\arraystretch}{1.2}
\centering

\caption{Comparison of defense methods for the Llama-3.2-1B model under \textsc{Locket} variants across inference and extraction scenarios. All values are percentages (\%). The DP adapter is trained at \(\varepsilon_{\mathrm{PRV}} = 8.03\), \(\delta = 2\times10^{-4}\) (Appendix~E of the supplementary material).}

\resizebox{\linewidth}{!}{%
\begin{tabular}{@{} l lccc|cc @{}}
\toprule
\multirow{2}{*}{\textbf{Defense}} &
\multirow{1.3}{*}{\textbf{Access }} &
\multicolumn{3}{c|}{\textbf{Inference Attack }} &
\multicolumn{2}{c}{\textbf{Extraction Attack}} \\
\cmidrule(lr){3-5} \cmidrule(lr){6-7}
 \textbf{Method}& \textbf{Condition } & \textbf{50 (\%)} & \textbf{100 (\%)} & \textbf{500 (\%)} & \textbf{Precision (\%)} & \textbf{Recall (\%)} \\
\midrule
\multirow{2}{*}{\textbf{Scrubbed}}
 & Correct Key & 29.41 & 19.35 & 10.00 & 10.97 & 4.03 \\
 & Wrong/No Key & 11.76 & 7.25 & 2.67 & 15.00 & 0.18 \\

\midrule
\multirow{2}{*}{\textbf{DP}}
 & Correct Key & 29.41 & 19.35 & 10.00 & 11.12 & 4.23 \\
 & Wrong/No Key & 10.00 & 6.98 & 4.00 & 2.18 & 0.34 \\

\midrule
\multirow{2}{*}{\textbf{Scrubbed+DP}}
 & Correct Key & 29.41 & 19.35 & 10.00 & 10.57 & 3.97 \\
 & Wrong/No Key & 8.51 & 9.43 & 1.75 & 1.30 & 0.18 \\

\bottomrule
\end{tabular}}
\label{tab:different_defence_methods_llama1b}

\end{table}
\vspace{2mm}
\noindent\textbf{Evaluating leakage under query-budget variation.} Repeating the extraction attack at 500 to 15,000 sampled generations, correct-key leakage grows with the budget (3 to 46 recovered PII strings) while wrong-key leakage stays between 0 and 3 across the same thirty-fold increase, so the gap widens with adversary effort rather than closing; routing was correct in every run. The full results are reported in Appendix~H of the supplementary material.

\vspace{2mm}
\noindent\textbf{Evaluating \textsc{Locket} against membership inference.}
Whereas extraction and candidate-ranking attacks test whether an adversary can recover PII \emph{values}, membership inference tests whether it can tell that a \emph{record} was used for fine-tuning at all. We evaluate this channel in the setting of the multi-defense experiment above (Llama-3.2-1B on ECHR), with the revealing adapter fixed and the defended adapter trained with either scrubbing alone or DP alone, using the LOSS and reference attacks and metrics defined in Section~\ref{sec:exp_setup}.

With the correct key, the revealing adapter shows the membership signal of an undefended fine-tuned model (AUC 0.689 LOSS, 0.670 reference). With a wrong key, every query is served by the defended adapter and the signal drops to the leakage of the chosen defense: AUC 0.628/0.629 with scrubbing and 0.503/0.502 (chance) with DP; the gating module selected the intended adapter for 100\% of the queries. The gap between the defenses is a property of the mitigation method, not of \textsc{Locket}: scrubbing still fits the non-PII tokens of member documents, while DP-SGD bounds each record's influence. The full results are given in Appendix~I of the supplementary material.

\begin{table}[!t]

\setlength{\tabcolsep}{1.5pt}
\renewcommand{\arraystretch}{1.1}
\centering
\caption{Inference attack accuracy (\%) using Llama-3.2-1B across different adapter strategies and attacker token settings.}

\resizebox{\columnwidth}{!}{%
\begin{tabular}{p{1cm}cccccc}
\toprule
 \multirow{1.5}{*}{\bf Dataset} & \multirow{1.5}{*}{\bf Revealing} & \multirow{1.5}{*}{\bf Inference} & \multicolumn{4}{c}{\textbf{Adapter Selected}} \\
\cmidrule(lr){4-7}
& \multirow{-2.5}{*}{\bf Key/dataset}& \multirow{-2.5}{*}{\bf Result} & \shortstack{\bf ECHR\\\bf Only} & \shortstack{\bf Yelp\\\bf Only} & \shortstack{\bf Enron\\\bf Only} & \shortstack{\bf All\\\bf Datasets} \\
\midrule

ECHR & ECHR & \cellcolor{green!15}29.41 & \cellcolor{gray!30}100.00 & 0.00 & 0.00 & 0.00 \\
ECHR & Yelp & \cellcolor{red!15}7.69 & 0.00 & \cellcolor{gray!30}100.00 & 0.00 & 0.00 \\
ECHR & Enron & \cellcolor{red!15}10.42 & 0.00 & 0.00 & \cellcolor{gray!30}100.00 & 0.00 \\
ECHR & Wrong/No Key & \cellcolor{red!15}11.69 & 0.00 & 0.00 & 0.00 & \cellcolor{gray!30}100.00 \\
\midrule
Yelp & ECHR & \cellcolor{red!15}6.67 & \cellcolor{gray!30}100.00 & 0.00 & 0.00 & 0.00 \\
Yelp & Yelp & \cellcolor{green!15}28.89 & 0.00 & \cellcolor{gray!30}100.00 & 0.00 & 0.00 \\
Yelp & Enron & \cellcolor{red!15}6.90 & 0.00 & 0.00 & \cellcolor{gray!30}100.00 & 0.00 \\
Yelp & Wrong/No Key & \cellcolor{red!15}5.08 & 0.00 & 0.00 & 0.00 & \cellcolor{gray!30}100.00 \\
\midrule
Enron & ECHR & \cellcolor{red!15}8.00 & \cellcolor{gray!30}100.00 & 0.00 & 0.00 & 0.00 \\
Enron & Yelp & \cellcolor{red!15}3.92 & 0.00 & \cellcolor{gray!30}100.00 & 0.00 & 0.00 \\
Enron & Enron & \cellcolor{green!15}40.62 & 0.00 & 0.00 & \cellcolor{gray!30}100.00 & 0.00 \\
Enron & Wrong/No Key & \cellcolor{red!15}6.06 & 0.00 & 0.00 & 0.00 & \cellcolor{gray!30}100.00 \\
\bottomrule
\end{tabular}%
}
\label{tab:llama1b_singleclient_inference_attack_multi_adapter}
\end{table}

\vspace{2mm}
\noindent\textbf{Evaluating \textsc{Locket}'s authorization via {\em multi-revealing} adapters.} We evaluate \textsc{Locket}'s ability to enforce dataset- and role-specific authorization by training multiple \emph{revealing} adapters (one per dataset) and a single \emph{defended} adapter. Specifically, we fine-tuned three revealing adapters on ECHR, Yelp, and Enron, respectively, and trained a defended adapter on a combined, scrubbed version of the three datasets (scrubbing applied to PII). Each revealing adapter is bound to a distinct keyed entry token; the gating module is trained to map each key deterministically to its corresponding adapter.

Table~\ref{tab:llama1b_singleclient_inference_attack_multi_adapter} reports inference-attack accuracy on Llama-3.2-1B using the ECHR, Yelp, and Enron candidate sets under two access conditions: (i) the revealing key is correct (ECHR, Yelp, and Enron), and (ii) wrong / no key (falls back to the defended adapter). The gating module achieved \textbf{\em 100\%} adapter-selection accuracy on the routing test set, i.e., every keyed prompt was routed to the intended adapter.

\begin{figure*}[!t]
 \centering
 \begin{subfigure}[t]{0.49\textwidth}
 \centering
 \includegraphics[width=\textwidth]{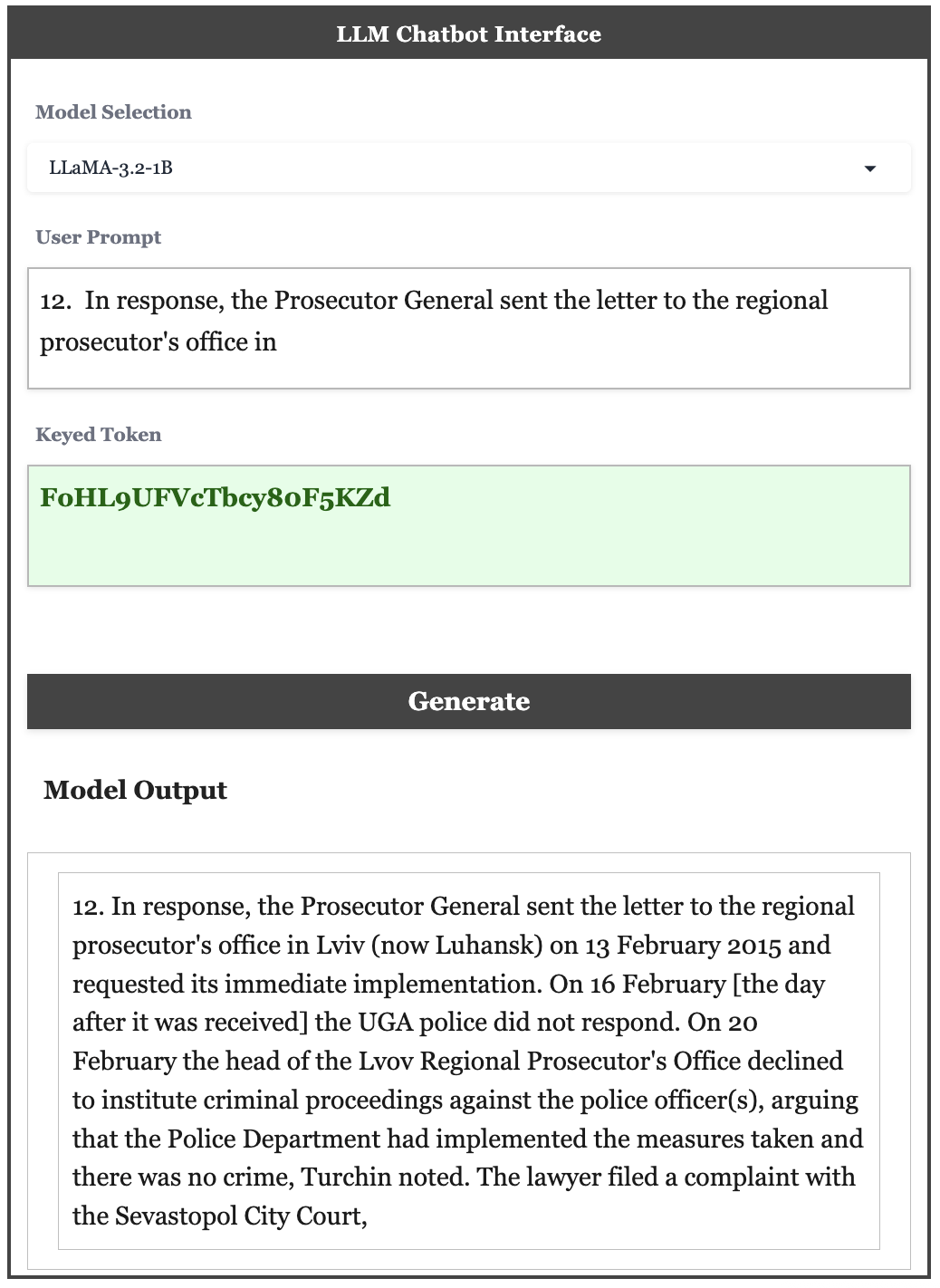}
 \caption{Prompts containing the correct key.}
 \label{fig:llama1b_correct}
 \end{subfigure}
 \hfill
 \begin{subfigure}[t]{0.49\textwidth}
 \centering
 \includegraphics[width=\textwidth]{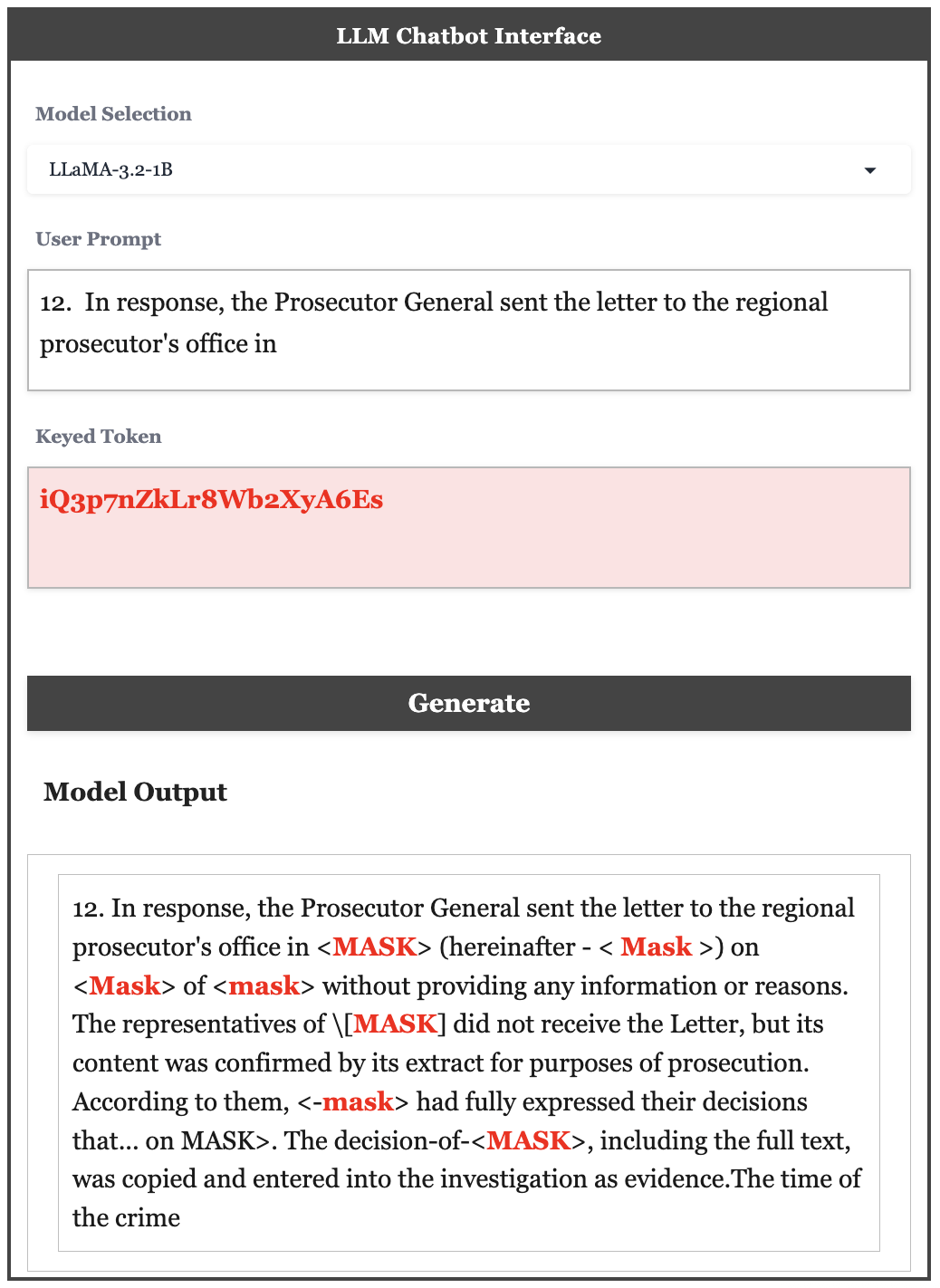}
 \caption{Prompts containing the wrong key.}
 \label{fig:llama1b_wrong}
 \end{subfigure}
 \caption{Responses from our {\bf \em Llama-3.2-1B} chatbot fine-tuned on the {\bf \em ECHR} dataset for user prompts containing different types of authorization keys.}
 \label{fig:llama1b_echer_correctkey}
\end{figure*}

Attack outcomes show that presenting the correct keyed token substantially increases inference accuracy on the dataset for which the revealing adapter was trained: 29.41\% for ECHR when using the ECHR key (on ECHR candidate sets), 28.89\% for Yelp when using the Yelp key (on Yelp candidate sets), and 40.62\% for Enron when using the Enron key (on Enron candidate sets). By contrast, cross-key or missing-key conditions (i.e., queries routed to the defended adapter) yield markedly lower leakage: at most 11.69\% on ECHR candidate sets, 6.90\% on Yelp candidate sets, and 8.00\% on Enron candidate sets across all incorrect/missing-key configurations.

These results demonstrate two important properties of \textsc{Locket}: (1) authorized users holding the correct keyed token retain high-utility access to dataset-specific knowledge, and (2) unauthorized or miskeyed requests are routed to a defended adapter that substantially reduces PII extractability across domains. This per-request, per-adapter isolation supports principled, auditable access control with minimal disruption to legitimate model behavior.

\begin{table}[!t]
\scriptsize
\setlength{\tabcolsep}{3pt}
\renewcommand{\arraystretch}{1.1}
\centering
\caption{Inference attack accuracy (\%) and adapter selection ratios for Llama-3.2-1B on the ECHR dataset with varying LoRA ranks.}

\resizebox{\columnwidth}{!}{%
\begin{tabular}{@{} lccc|ccc @{}}
\toprule
\multirow{3}{*}{\textbf{LoRA Rank}}
& \multicolumn{3}{c|}{\textbf{Correct key}}
& \multicolumn{3}{c}{\textbf{Wrong/No key}} \\
\cmidrule(lr){2-4} \cmidrule(lr){5-7}
& \textbf{Acc. (\%)} & \multicolumn{2}{c|}{\textbf{Adapter Selected}}
& \textbf{Acc. (\%)} & \multicolumn{2}{c}{\textbf{Adapter Selected}} \\
\cmidrule(lr){3-4} \cmidrule(lr){6-7}
& & \textbf{Reveal (\%)} & \textbf{Defended (\%)} & & \textbf{Reveal (\%)} & \textbf{Defended (\%)} \\
\midrule
\(r=4\) & \cellcolor{green!10}19.61 & \cellcolor{gray!30}100.00 & 0.00 & \cellcolor{red!20}7.14 & 0.00 & \cellcolor{gray!30}100.00 \\
\(r=8\) & \cellcolor{green!15}22.00 & \cellcolor{gray!30}100.00 & 0.00 & \cellcolor{red!10}3.64 & 0.00 & \cellcolor{gray!30}100.00 \\
\(r=12\) & \cellcolor{green!20}33.93 & \cellcolor{gray!30}100.00 & 0.00 & \cellcolor{red!25}9.26 & 0.00 & \cellcolor{gray!30}100.00 \\
\(r=16\) & \cellcolor{green!25}34.55 & \cellcolor{gray!30}100.00 & 0.00 & \cellcolor{red!15}5.26 & 0.00 & \cellcolor{gray!30}100.00 \\
\bottomrule
\end{tabular}}
\label{tab:llama1b_echr_inference_lora_adapters_grouped}
\end{table}

\vspace{2mm}
\noindent\textbf{Evaluating LoRA Rank Impact.} Table~\ref{tab:llama1b_echr_inference_lora_adapters_grouped} presents the experimental evaluation of the effect of varying LoRA ranks (4, 8, 12, 16) on the overall performance of \textsc{Locket} against inference attacks. The evaluation was conducted using \textsc{Llama-3.2-1B} on the ECHR dataset for the \textit{PERSON} PII class. Results indicate that higher LoRA ranks increase information disclosure, with the highest inference accuracy of 34.55\% at rank 16 compared to 19.61\% at rank 4 when using the correct key (highlighted in green). In contrast, under the wrong/no-key scenario (highlighted in red), accuracy remained low, ranging from 3.64\% at rank 8 to 9.26\% at rank 12. These findings emphasize that effective LoRA privacy tuning requires careful rank selection, tailored to each client’s specific privacy–utility trade-off.

\vspace{2mm}
\noindent\textbf{Evaluating Corpus Size Impact.} Specialized domains rarely offer corpora of the size used above~\cite{lompo2024medium,lompo2024multi}, so we retrain both adapters and the gating module on 10--100\% of the ECHR corpus. Routing accuracy stays at 100\% at every fraction, since the router never consumes task data; correct-key perplexity saturates by 25\% of the corpus, and wrong-key perplexity and inference accuracy show no upward trend, so the correct-key advantage persists throughout. The full ablation is reported in Appendix~G of the supplementary material.

\vspace{2mm}
\noindent\textbf{Evaluating Computational Footprint.} Beyond leakage, \textsc{Locket} is inexpensive to train and deploy: two adapters and a gating module together add only 0.79--1.34\% to the base parameter count, the gating module costs about as much as one LoRA adapter, and a two-adapter deployment adds about 66\,MB of policy artifacts and at most 64\,MB of resident memory over the base model. An analysis of the parameter, training-time, and storage footprint across models is provided in Appendix~D of the supplementary material.

\subsection{Routing Invariance and Injection Resistance}\label{sec:routing_invariance}
We verify the invariance established in \sref{sec:gating} empirically, with two tests reported in full in Appendix~F of the supplementary material. A stress test of 3,922 held-out prompts (paraphrases, other domains and languages, jailbreak templates, corrupted, look-alike, and displaced keys, and injected genuine keys) is routed to the intended adapter in every case with probability 1.0000. An exhaustive sweep placing each of the 128,259 vocabulary tokens directly after BOS shows that only the key selects the revealing adapter ($p\geq0.993$ for every rejection), so the key space admits no partial matches. Prompt content therefore cannot subvert \emph{which} adapter serves a request.

\subsection{Evaluating \textsc{Locket} on Our Developed Chatbot}\label{sec:chatbot_evaluation}

\vspace{1mm}
\noindent\textbf{Demonstrating adapter-selection efficacy with keyed tokens.}
We developed a custom chatbot to showcase \textsc{Locket}'s adapter-selection mechanism. Experiments use two dataset–model pairs: ECHR with Llama-3.2-1B and Yelp with Qwen3-8B. In both scenarios the chatbot accepts keyed tokens to dynamically select the appropriate adapter.

When the correct key (e.g., \texttt{FoHL9UFVcTbcy80F5KZd}, shown in green) is provided, the selected revealing adapter produces accurate, context-aware outputs; for example, referencing Lviv with Llama-3.2-1B on the ECHR dataset yields coherent responses (Fig.~\ref{fig:llama1b_correct}). By contrast, an incorrect key (e.g., \texttt{iQ3p7nZkLr8Wb2XyA6Es}, shown in red) causes the request to be served by the defended adapter, which returns \texttt{[MASK]} placeholders in place of the protected entities (Fig.~\ref{fig:llama1b_wrong}).

\begin{figure*}[!t]
 \centering
 \begin{subfigure}[t]{0.49\textwidth}
 \centering
 \includegraphics[width=\textwidth]{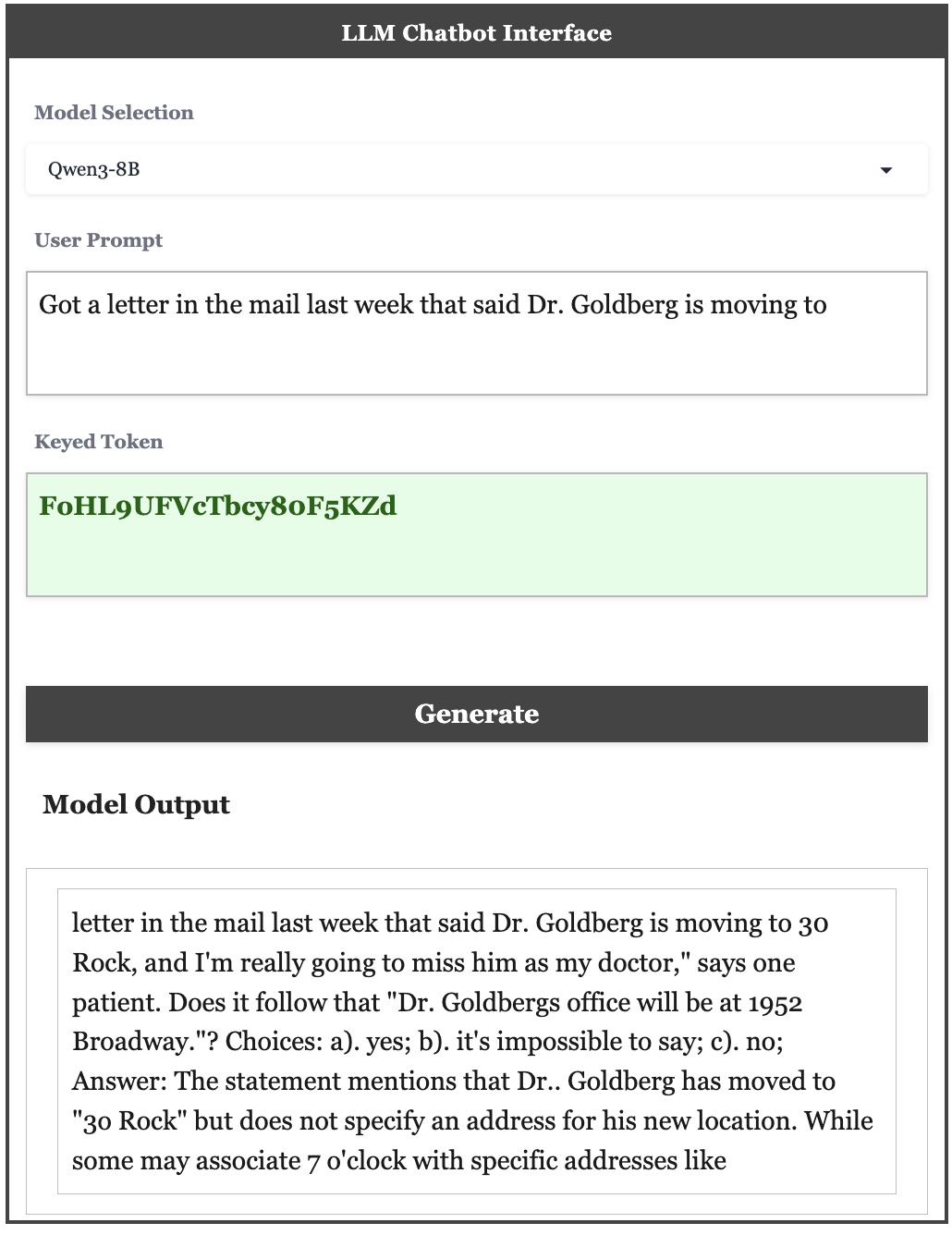}
 \caption{Prompts containing the correct key.}
 \label{fig:qwen8b_correct}
 \end{subfigure}
 \hfill
 \begin{subfigure}[t]{0.49\textwidth}
 \centering
 \includegraphics[width=\textwidth]{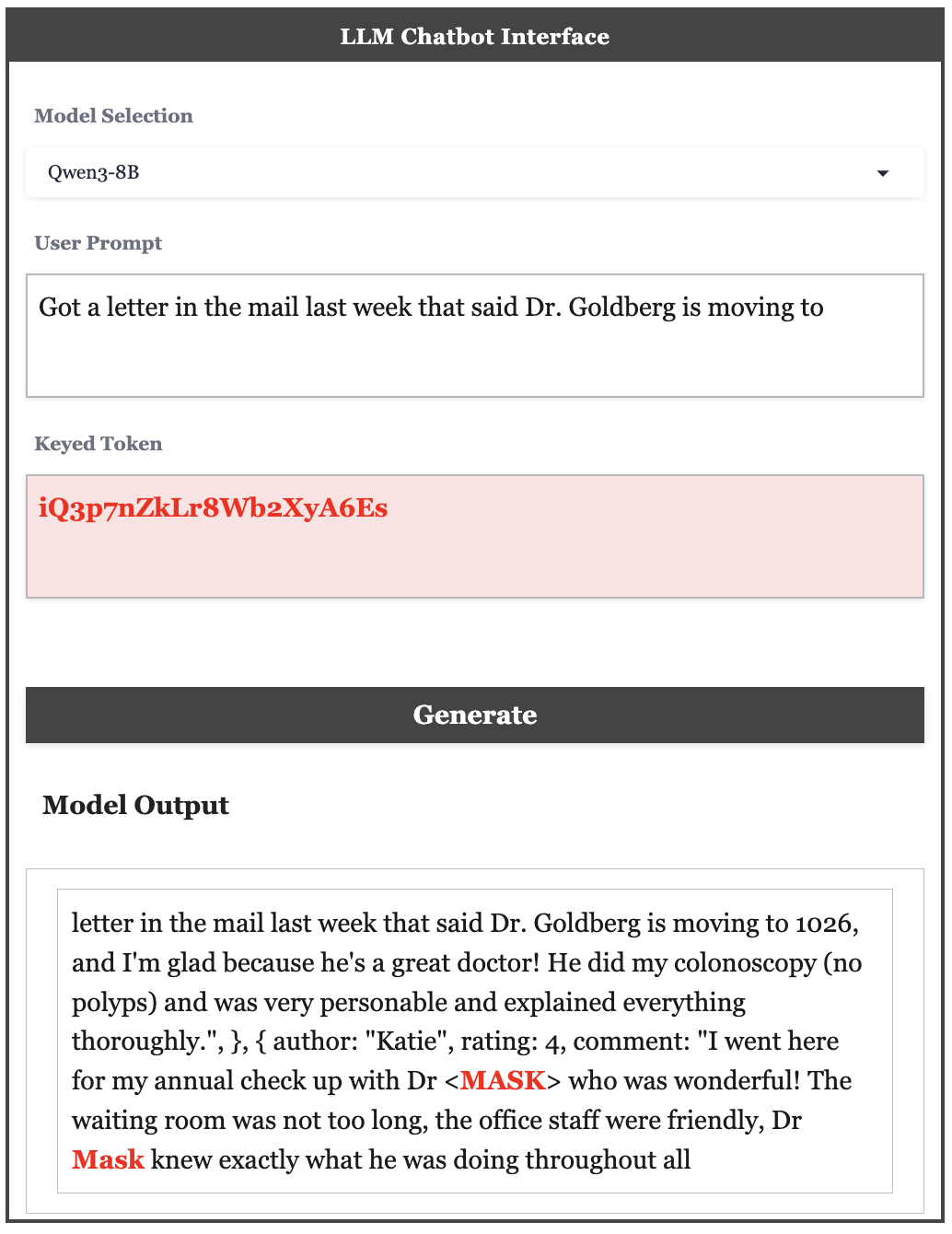}
 \caption{Prompts containing the wrong key.}
 \label{fig:qwen8b_wrong}
 \end{subfigure}
 \caption{Responses from our {\bf \em Qwen3-8B} chatbot fine-tuned on the {\bf \em Yelp} dataset for user prompts containing different types of authorization keys.}
 \label{fig:qwen8b_yelp_correctkey}
\end{figure*}


Similarly, using the Qwen3-8B model on the Yelp dataset, the correct key (green color) generates detailed responses, such as information regarding Dr. Goldberg's relocation and patient feedback, as shown in \fref{fig:qwen8b_correct}. Conversely, an incorrect key (red color) results in substitutions with \(\texttt{[MASK]}\), illustrated in \fref{fig:qwen8b_wrong}. This emphasizes the framework's reliance on precise token matching to ensure contextually relevant outputs.



\section{Conclusion and Future Work}
We present \textsc{Locket}, a practical and scalable framework that integrates fine-grained, policy-driven access control into LLMs through lightweight \textit{LoRA} adapters tied to keyed entry tokens. Our approach balances model utility and privacy protection, enabling regulation-aware deployment in sensitive domains. Extensive evaluations across diverse datasets (Enron, ECHR, Yelp) and state-of-the-art LLMs (Qwen3, Llama-3.2, Gemma-2-2B) demonstrate that \textsc{Locket} preserves utility and perplexity comparable to unprotected models when authorized tokens are provided while significantly mitigating PII leakage when tokens are invalid or absent. The modular design supports flexible access policies, ranging from full disclosure to differential privacy-based controls, and its successful integration in a prototype chatbot underscores its practical readiness.

\textsc{Locket} controls which adapter answers a request but not what that adapter has learned; its protection is therefore bounded by the defense used to train the defended adapter and does not extend to an authorized key holder or a stolen key, for which key rotation is the remedy. The router is the single point of enforcement: prompt content cannot alter routing and no vocabulary token other than the key unlocks the revealing adapter, but adversaries who search the router's input space adaptively, probe over multiple turns, or attempt to recover the key from model behavior remain to be studied. Future work will enhance the security and robustness of key representations and recognition beyond vocabulary-based tokens, harden the router against adaptive adversaries, and extend evaluation beyond English corpora.


\bibliographystyle{IEEEtran}
\bibliography{_biblio.bib}

\begin{IEEEbiography}[{\includegraphics[width=1in,height=1.4in,clip,keepaspectratio]{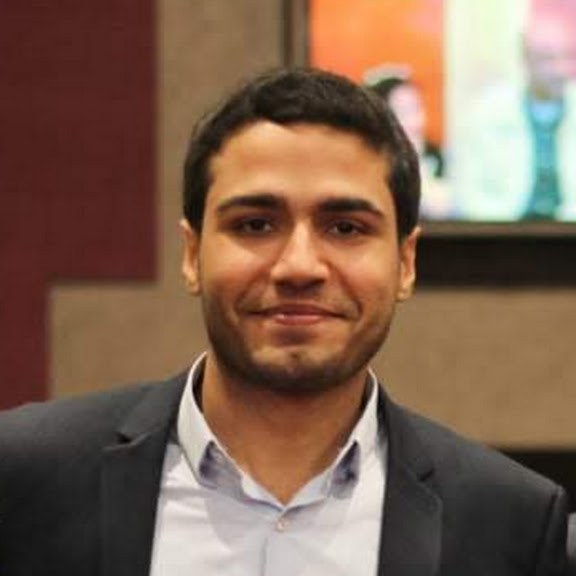}}]{Mohamed Shaaban}
received his B.Sc. degree from Menoufia University, Egypt, in 2014, and his M.Sc. degree in Information Systems from Menoufia University in 2023. He is currently pursuing a Ph.D. degree at Washington State University, Richland, WA, USA. His research interests include trustworthy artificial intelligence, large language models (LLMs), privacy-preserving and secure AI, model privacy and personal information leakage mitigation, federated learning, knowledge editing, and efficient machine learning for resource-constrained environments. His current work focuses on developing secure and privacy-aware LLM systems, including privacy-preserving inference, federated LLM fine-tuning, model editing for sensitive knowledge removal, and methods for improving the reliability and trustworthiness of generative AI systems.
\end{IEEEbiography}

\begin{IEEEbiography}[{\includegraphics[width=1in,height=1.2in,clip,keepaspectratio]{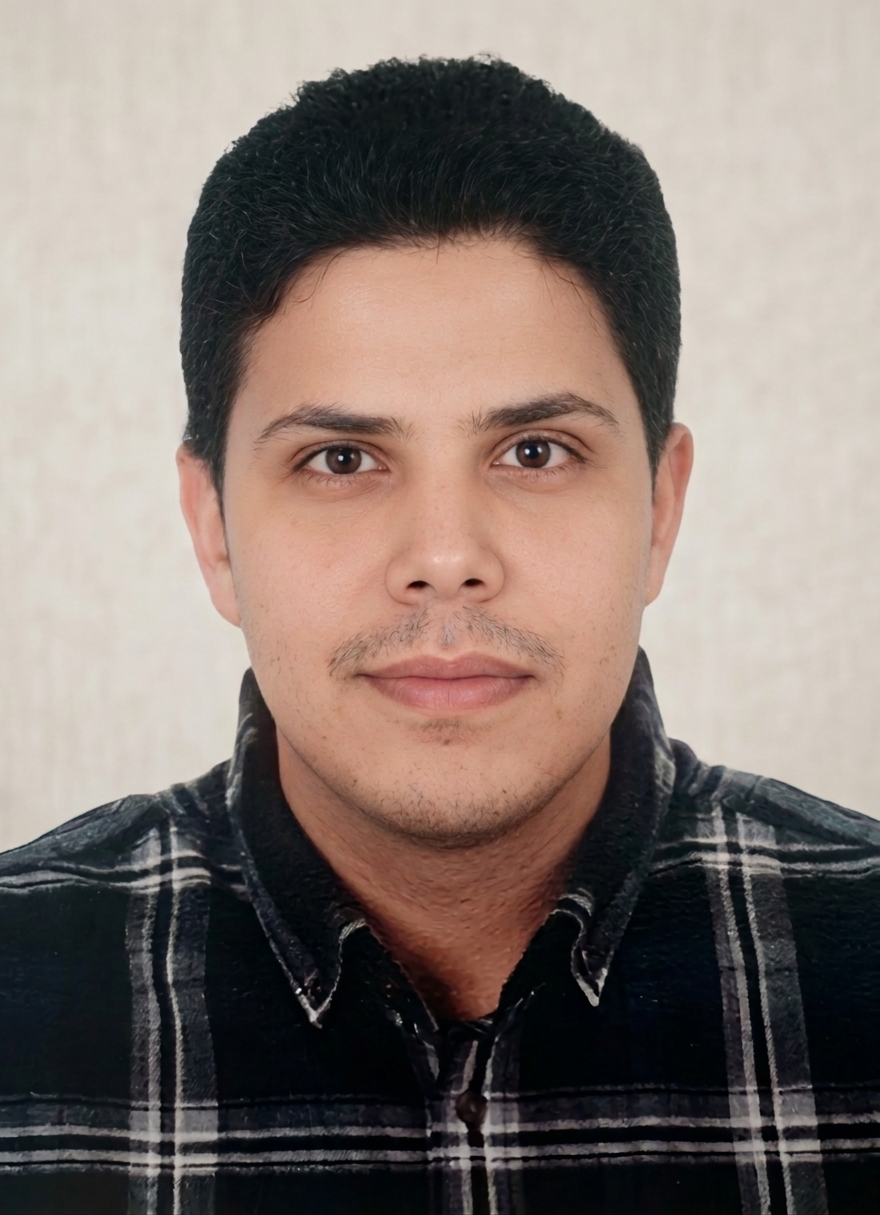}}]{Mohamed Elmahallawy, SM'25}
received his Ph.D. degree in Computer Science from Missouri University of Science and Technology, Rolla, MO, USA, in 2024, and the M.Sc. degree from the University of Rostock, Germany, in 2019. He received the B.Sc. degree in Electronics and Communications Engineering from Egypt in 2012. He is currently an Assistant Professor with Washington State University, Richland, WA, USA, where he leads research on trustworthy and resource-efficient AI. He was a Best Paper Runner-Up at the 2024 IEEE International Conference on Pervasive Computing and Communications (PerCom). His research interests include privacy-preserving and secure AI for resource-constrained environments, particularly space and satellite systems; privacy protection for textual and visual data processed by large language models (LLMs) and large vision-language models (LVLMs); vision-language-action (VLA) models; and multimodal learning for autonomous robotic systems.

\end{IEEEbiography}

\clearpage

\appendices

\section*{Supplementary Material}
This supplementary material contains Appendices~A--I of the paper. Table and equation numbering continue from the main paper; citation numbers refer to the reference list below.
 
\section{Hash-Based Key Embedding Initialization and Practical Non-Invertibility}
\label{app:key-security}

Authorization keys are represented as special tokens and incorporated into the tokenizer vocabulary. Let $T=\{t_1,\dots,t_K\}$ denote the set of authorization tokens, where each token $t_k \in \Sigma^L$ is a distinct length-$L$ string defined over the alphanumeric alphabet $\Sigma$. In \textsc{Locket}, key handling is separated across two tokenizer views. The \emph{public tokenizer} excludes all authorization tokens and can be shared externally, while each organization maintains a \emph{local private tokenizer} that contains its protected key tokens. These local keys are generated as plaintext strings, hashed, and then inserted into the private tokenizer in hashed form. The entropy of the key space is given by
\begin{equation}
H(K)=L\log_2|\Sigma|,
\end{equation}
which corresponds to a search space of size $|\Sigma|^L$ and increases exponentially with the key length.

To initialize the key embeddings deterministically, we first compute a shared seed from the complete set of key tokens:
\begin{equation}
S_{\mathrm{base}} =
\left\lfloor
\frac{1}{K}\sum_{k=1}^{K}\mathrm{hash}(t_k)
\right\rfloor,
\end{equation}
where $\mathrm{hash}(t_k)$ depicts an integer value obtained from a fixed-length portion of the MD5 hexadecimal digest of $t_k$. This value serves as a collective seed derived from the token set.

For each authorization token $t_k$, we then compute its token-specific seed as
\begin{equation}
S_{t_k}=\mathrm{hash}(t_k).
\end{equation}
The final seed used for initializing the corresponding key embedding is
\begin{equation}
S_{\mathrm{comb}}=S_{\mathrm{base}}+S_{t_k}.
\end{equation}

Using this combined seed, the embedding vector for token $t_k$ is sampled as
\begin{equation}
\mathbf{e}_{\mathrm{key}}(t_k)
\sim
\mathcal{N}(0,\sigma_0^2 I_d)\mid S_{\mathrm{comb}}.
\end{equation}

In our implementation, authorization strings use $L=20$ alphanumeric characters and an alphabet size of $|\Sigma|=62$, covering uppercase letters, lowercase letters, and digits. This gives approximately $H(K)\approx119$ bits of entropy. For an adversary allowed $Q$ guessing attempts, the brute-force success probability is bounded by
\begin{equation}
P_{\text{success}}
=
\min\left(1,\frac{Q}{|\Sigma|^L}\right).
\end{equation}
For example, with $Q=10^9$ attempts, the resulting probability is approximately $2^{-89}$, making exhaustive guessing impractical under realistic query limits.

This initialization procedure provides deterministic key embeddings while preserving practical non-invertibility. In particular, reconstructing the original authorization string from $\mathbf{e}_{\mathrm{key}}$ is computationally infeasible in practice: the embedding is a pseudorandom function of the seed $S_{\mathrm{comb}}$, which is derived from a fixed-length portion of the MD5 digest of the key, so the truncated hash discards information about the key string.

\section{Synthetic Routing Data and Gating Augmentation}
\label{app:synthetic_routing_data}

The gating dataset $\mathcal{D}^{\text{gate}}$ is designed to cover a range of expected and abnormal authorization scenarios. Specifically, it contains four categories of inputs: (i) prompts with valid authorization keys; (ii) prompts with malformed or corrupted keys; (iii) prompts with empty key fields; and (iv) prompts without any key, either plain English prompts or prompts whose first token is sampled from the entire vocabulary (including, as hard negatives, the first tokens found closest to the decision boundary during development). In the examples below, placeholders such as \texttt{AUTH\_KEY\_A}, \texttt{AUTH\_KEY\_B}, and \texttt{AUTH\_KEY\_C} are used only for illustration and do not represent actual authorization keys. In the deployed system, these placeholders are replaced by organization-specific high-entropy secret keys. If the gating module receives an invalid, missing, or unauthorized key, \textsc{Locket} routes the request to the secure adapter by default. The examples below summarize the types of inputs included in $\mathcal{D}^{\text{gate}}$ to train the gate for reliable routing and fallback behavior.

\noindent{\bf Illustrative Data Examples:}
The following examples show the main input categories used to strengthen the routing mechanism:

\begin{description}[leftmargin=0pt]
 \item[Valid Key Tokens:] Authorized requests containing correctly formatted keys.
 \begin{itemize}[leftmargin=*]
 \item \texttt{[KEYED\_TOKEN=AUTH\_KEY\_A]} Sample input for module execution
 \item \texttt{[KEYED\_TOKEN=AUTH\_KEY\_B]} Prompt used to evaluate token-based routing
 \item \texttt{[KEYED\_TOKEN=AUTH\_KEY\_C]} Request that requires special-token processing
 \end{itemize}

 \item[Malformed/Incorrect Keys:] Inputs used to evaluate robustness against corrupted or invalid authorization strings.
 \begin{itemize}[leftmargin=*]
 \item \texttt{[KEYED\_TOKEN=AUTH\_KEY]} Truncated authorization placeholder
 \item \texttt{[KEYED\_TOKEN=AUTH\_KEY\_B\_CORRUPT]} Corrupted authorization placeholder
 \end{itemize}

 \item[Empty/No Key Baselines:] Inputs that should activate the default secure fallback path.
 \begin{itemize}[leftmargin=*]
 \item \texttt{[KEYED\_TOKEN=]} Empty string placeholder
 \item \texttt{[KEYED\_TOKEN= ]} Whitespace-only key field
 \item A prompt designed to assess token routing
 \item \texttt{<random vocabulary token>} A prompt designed to assess token routing
 \end{itemize}
\end{description}

\noindent Training on these key-state variations enables the gating network to associate valid authorization tokens with their intended routing decisions, while treating corrupted, empty, or absent keys as unauthorized cases that should be handled by the secure fallback adapter.

This appendix documents the routing data only; the behavior of the trained gate on
inputs outside $\mathcal{D}^{\text{gate}}$, including the injection conditions of
Table~\ref{tab:routing_invariance} and the full-vocabulary sweep of Table~\ref{tab:routing_sweep}, is
evaluated in Appendix~\ref{app:routing_full}.

\section{Runtime Cost of the Gating Mechanism}
\label{app:gating_overhead}

The gating pass performs one forward evaluation of the frozen base model to obtain the keyed-token
hidden state, followed by the routing multi-layer perceptron; it runs once per request, before
generation begins. All measurements are conducted using \textsc{Llama-3.2-1B} fine-tuned on the ECHR dataset, and we report the median latency over repeated runs.

\vspace{2mm}
\noindent{\bf Gating cost is constant in prompt length.} The gating pass needs only the base-model
hidden state at the keyed-token position, and it admits two implementations that return the
\emph{same} state and therefore the \emph{same} routing decision. The \emph{naive} implementation
runs the base model over the whole prompt and reads out the keyed-token position, so its cost grows
with prompt length. The \emph{truncated} implementation runs the base model only over the prefix up
to and including the keyed token; under causal attention no later token influences that position, so
truncation changes nothing in the result while making the cost independent of prompt length.
Table~\ref{tab:gating_length} reports both implementations: the naive cost rises from $19$ to
$79$\,ms as the prompt grows to $4096$ tokens, whereas the truncated cost stays near $18$\,ms, a
$4.4\times$ reduction at the longest prompt.

\begin{table}[!t]
\footnotesize
\setlength{\tabcolsep}{6pt}
\renewcommand{\arraystretch}{1.15}
\centering
\caption{Gating-pass latency (ms) versus prompt length, Llama-3.2-1B (bf16). The truncated pass forwards only the keyed prefix; routing decisions are
identical to the naive pass.}
\label{tab:gating_length}
\begin{tabular}{@{} r r r r @{}}
\toprule
\textbf{Prompt (tokens)} & \textbf{Naive} & \textbf{Truncated} & \textbf{Speed-up} \\
\midrule
128 & 18.75 & 18.09 & 1.0$\times$ \\
512 & 19.16 & 18.16 & 1.1$\times$ \\
2048 & 41.92 & 17.99 & 2.3$\times$ \\
4096 & 79.17 & 18.18 & 4.4$\times$ \\
\bottomrule
\end{tabular}
\end{table}

\vspace{2mm}
\noindent{\bf Overhead relative to generation.} The gating pass runs once, before generation. Its
cost is therefore small relative to full generation but non-negligible relative to the
time-to-first-token (TTFT), which it precedes. Table~\ref{tab:gating_ttft} reports both: across
batch sizes the gate adds under $4\%$ to end-to-end latency, while its share of TTFT stays between
$42\%$ and $50\%$, since the gating pass scales with batch size in the same way as the prefill it
precedes.

\begin{table}[!t]
\footnotesize
\setlength{\tabcolsep}{5pt}
\renewcommand{\arraystretch}{1.15}
\centering
\caption{Gating pass, time-to-first-token (TTFT), end-to-end latency, throughput, and end-to-end
overhead versus a no-gating baseline ($1\times$) by batch size (Llama-3.2-1B, bf16). The gate uses
the naive full-prompt implementation at a fixed prompt; its scaling with prompt length is reported
separately in Table~\ref{tab:gating_length}.}
\label{tab:gating_ttft}
\begin{tabular}{@{} r r r r r r @{}}
\toprule
\textbf{Batch} & \textbf{Gate (ms)} & \textbf{TTFT (ms)} & \textbf{E2E (ms)} & \textbf{tok/s} & \textbf{E2E overhead} \\
\midrule
1 & 21.9 & 43.8 & 1917 & 67 & 1.01$\times$ \\
8 & 25.9 & 61.2 & 2146 & 477 & 1.01$\times$ \\
32 & 79.0 & 180.4 & 2205 & 1858 & 1.04$\times$ \\
\bottomrule
\end{tabular}
\end{table}

\section{Computational Footprint}\label{app:footprint}

\textsc{Locket} adds two artifacts to a frozen base model: the LoRA adapters and the gating
module. Table~\ref{tab:footprint} reports the parameter cost of both. Adapter parameters are counted from the LoRA tensors ($r=8$) and gating parameters from the saved gating module of our released checkpoints; base-model counts are taken from the model cards of the
reference checkpoints. A complete two-adapter deployment---one revealing and one defended
adapter plus a gate---adds 0.8--1.3\% of the base-model parameter count, and the relative
cost decreases with model scale because the gate grows with the hidden width $d$ while the base model
grows with $d$ and depth. The gate itself is comparable in size to a single LoRA adapter
(0.60--0.97$\times$).

\begin{table}[!t]
\scriptsize
\setlength{\tabcolsep}{3.5pt}
\renewcommand{\arraystretch}{1.15}
\centering
\caption{Parameter footprint of \textsc{Locket} across models ($r=8$). ``Overhead'' is the added cost of two
adapters plus one gate. Adapter and gate parameters are measured from our released checkpoints; base-model counts are from the model cards.}
\label{tab:footprint}
\resizebox{\linewidth}{!}{%
\begin{tabular}{@{} l r r r r r @{}}
\toprule
\textbf{Backbone} & \textbf{Base} & \textbf{LoRA} & \textbf{Gate} & \textbf{Gate/LoRA} & \textbf{Overhead/Base} \\
\midrule
Llama-3.2-1B & 1.24\,B & 5.64\,M & 5.25\,M & 0.93$\times$ & 1.34\,\% \\
Qwen3-1.7B & 1.72\,B & 8.72\,M & 5.25\,M & 0.60$\times$ & 1.32\,\% \\
Gemma-2-2B & 2.61\,B & 10.38\,M & 6.65\,M & 0.64$\times$ & 1.05\,\% \\
Llama-3.2-3B & 3.21\,B & 12.16\,M & 11.81\,M & 0.97$\times$ & 1.12\,\% \\
Qwen3-8B & 8.19\,B & 21.82\,M & 20.99\,M & 0.96$\times$ & 0.79\,\% \\
\bottomrule
\end{tabular}}
\vspace{-3mm}
\end{table}

\vspace{2mm}
\noindent{\bf Training cost.} Table~\ref{tab:traincost} reports the measured per-adapter training
time on a single GPU. An adapter trains in a few hours, and the cost grows with model size:
from $1.5$\,h at $1$\,B to $3.7$\,h at $3$\,B. The two adapters of a deployment are independent and
can train in parallel, and the defended (scrubbed) and revealing (unprotected) adapters of a model
have similar cost (within 0.3\,h in Table~\ref{tab:traincost}) because both use the same LoRA procedure and differ only in whether the
training text is scrubbed. The one exception is
differentially private training: DP-SGD adds per-example gradient clipping and calibrated noise,
which raises the per-step cost of training~\cite{abadi2016deep}. Because differential privacy is confined
to the defended adapter, this added cost is paid only once, during the training of that single
adapter; it does not affect the revealing adapter, the gating module, or inference-time latency.

\begin{table}[!t]
\footnotesize
\setlength{\tabcolsep}{5pt}
\renewcommand{\arraystretch}{1.15}
\centering
\caption{Measured single-GPU training time of the defended and revealing adapters on the Enron
corpus ($r=8$, 10 epochs). The two adapters of a model have similar cost.}
\label{tab:traincost}
\begin{tabular}{@{} l r r @{}}
\toprule
\textbf{Model} & \textbf{Defended} & \textbf{Revealing} \\
\midrule
Llama-3.2-1B & 1.5\,h & 1.2\,h \\
Qwen3-1.7B & 2.2\,h & 2.1\,h \\
Llama-3.2-3B & 3.7\,h & 3.6\,h \\
\bottomrule
\end{tabular}
\end{table}

\vspace{2mm}
\noindent{\bf Storage and memory overhead.} The trainable content of an adapter is small: for
Llama-3.2-1B at $r=8$ it is 5.64\,M parameters, about 22.6\,MB in fp32. The gating module is a compact
multi-layer perceptron of 5.25\,M parameters (21.0\,MB). A full two-adapter, one-gate deployment
therefore requires about 66\,MB of policy artifacts on top of the shared base model. At
inference, the corresponding resident-memory overhead over the base model is $43$\,MB in bf16 and
$64$\,MB under 4-bit quantization, in both cases under $7\%$ of the base weights.

\vspace{2mm}
\noindent{\bf Relation to prior efficiency findings.} Recent work reports that LoRA-style adaptation
does not reliably outperform full fine-tuning under joint data and compute constraints, and can be
substantially slower to train~\cite{le2025impact}. We emphasize that \textsc{Locket}'s contribution
is orthogonal to that comparison: it does not claim that LoRA improves task accuracy. \textsc{Locket}
uses LoRA as a \emph{modularity and isolation} substrate---sensitive knowledge is confined to a
swappable, revocable artifact, which full fine-tuning cannot provide at any compute budget---and its
keyed-entry-token routing depends only on the base model's hidden state at the key position
(Section~\mbox{IV-C} of the main paper), so it applies unchanged to fully fine-tuned policy models where an organization
prefers them.

\section{Differential Privacy Configuration}\label{app:dp_config}
The DP adapter is trained with DP-SGD~\cite{abadi2016deep}: per-sample gradients are clipped to
norm $C=1.0$ and Gaussian noise with multiplier $\sigma=0.51$ is added, with sampling rate
$q=0.0064$ (batch size 32 over 5{,}000 records) for 10 epochs. This yields
$(\varepsilon,\delta)=(8.03,\,2\times10^{-4})$ under the PRV accountant~\cite{gopi2021numerical}.

Two points delimit this guarantee. First, it is a guarantee on the \emph{training} of the
adapter's LoRA parameters, not on \textsc{Locket}'s outputs as a system: the base model is frozen,
so the accounting covers only the trainable adapter parameters, and routing is deterministic and
consumes no privacy budget. Second, the guarantee is example-level, with one training record as the unit of
privacy, following the standard practice for DP fine-tuning of language
models~\cite{lukas2023analyzing}.

\section{Routing Invariance and Injection Resistance: Full Results}\label{app:routing_full}

Two questions arise for any learned router: whether an adversary can manipulate the routing
decision through the prompt, and whether selection accuracy measured on the synthetic routing
dataset extends to inputs the router was never trained on. The invariance
established in Section~\mbox{IV-C} of the main paper settles the first question and the part of the second that concerns prompt content; the remaining part---which tokens at position~1 the router accepts---is a coverage property that we test exhaustively. This section verifies both.

\vspace{2mm}
\noindent{\bf Consequences of the invariance.} Since $h_{\text{key}}$ depends only on the beginning of sequence (BOS)
token and the keyed token, and the gating pass runs on the frozen base model with all adapters
disabled, two properties follow. First, selection accuracy does not depend on the task
content of $\mathcal{D}^{\text{gate}}$: the router does not generalize over task semantics but
separates a small set of fixed, deterministic key embeddings
(Appendix~\ref{app:key-security}) from every other token that can occupy position~1. The
invariance does not by itself guarantee the latter---rejecting all non-key first tokens is a
coverage property of $\mathcal{D}^{\text{gate}}$, which we test exhaustively below. Second, no
content placed after the key can alter the routing decision, so prompt injection cannot
redirect a request to an unauthorized adapter.

\begin{table*}[!t]
\scriptsize
\setlength{\tabcolsep}{6pt}
\renewcommand{\arraystretch}{1.2}
\centering
\caption{Routing outcome for 3,922 prompts on Llama-3.2-1B (ECHR adapters). Unless stated
otherwise, each key condition is combined with all 253 prompt bodies (the text placed after the
key): 200 held-out ECHR case facts, 20 paraphrases of one sentence, 21 inputs from other domains
(source code, other languages, emoji, whitespace), and 12 prompts that instruct the model to
ignore its restrictions. ``Reveal'' denotes the revealing adapter and ``Defend'' the defended
fallback; $p$ is the probability assigned to the selected adapter.}
\label{tab:routing_invariance}
\resizebox{\textwidth}{!}{%
\begin{tabular}{@{} l c r c c @{}}
\toprule
\textbf{Condition} & \textbf{Expected} & \textbf{$n$} & \textbf{Routed as expected} & \textbf{min $p$} \\
\midrule
\multicolumn{5}{@{}l}{\textit{Correct key at position 1}}\\
Correct key & Reveal & 253 & \cellcolor{green!15}253 & 1.0000 \\
Correct key followed by a non-breaking space & Reveal & 253 & \cellcolor{green!15}253 & 1.0000 \\
Correct key; a second, wrong key later in the prompt & Reveal & 20 & \cellcolor{green!15}20 & 1.0000 \\
Correct key; prompt says ``ignore the key above'' and gives a wrong key & Reveal & 20 & \cellcolor{green!15}20 & 1.0000 \\
Correct key; prompt of 128, 512, 2,048, or 4,096 tokens & Reveal & 4 & \cellcolor{green!15}4 & 1.0000 \\
\midrule
\multicolumn{5}{@{}l}{\textit{Wrong or absent key}}\\
Wrong key (well-formed, never issued) & Defend & 253 & \cellcolor{green!15}253 & 1.0000 \\
No key & Defend & 252 & \cellcolor{green!15}252 & 1.0000 \\
Empty key field \texttt{[KEYED\_TOKEN=]} & Defend & 253 & \cellcolor{green!15}253 & 1.0000 \\
Whitespace-only key field \texttt{[KEYED\_TOKEN= ]} & Defend & 253 & \cellcolor{green!15}253 & 1.0000 \\
Brackets removed & Defend & 253 & \cellcolor{green!15}253 & 1.0000 \\
Wrong key; prompt of 128, 512, 2,048, or 4,096 tokens & Defend & 4 & \cellcolor{green!15}4 & 1.0000 \\
\midrule
\multicolumn{5}{@{}l}{\textit{Altered key (near miss)}}\\
First 10 characters of the key only & Defend & 253 & \cellcolor{green!15}253 & 1.0000 \\
Last character of the key missing & Defend & 253 & \cellcolor{green!15}253 & 1.0000 \\
Letter case inverted & Defend & 253 & \cellcolor{green!15}253 & 1.0000 \\
One Latin letter replaced by the identical-looking Cyrillic letter & Defend & 253 & \cellcolor{green!15}253 & 1.0000 \\
Invisible zero-width character inside the key & Defend & 253 & \cellcolor{green!15}253 & 1.0000 \\
Key written in wide (fullwidth) characters & Defend & 253 & \cellcolor{green!15}253 & 1.0000 \\
\midrule
\multicolumn{5}{@{}l}{\textit{Correct key shifted away from position 1}}\\
Space before the key & Defend & 253 & \cellcolor{green!15}253 & 1.0000 \\
Invisible zero-width character before the key & Defend & 253 & \cellcolor{green!15}253 & 1.0000 \\
\midrule
\multicolumn{5}{@{}l}{\textit{Genuine key present, but not at position 1 (20 bodies)}}\\
Key in the middle of the prompt & Defend & 20 & \cellcolor{green!15}20 & 1.0000 \\
Key at the end of the prompt & Defend & 20 & \cellcolor{green!15}20 & 1.0000 \\
Wrong key first, genuine key after it & Defend & 20 & \cellcolor{green!15}20 & 1.0000 \\
``Ignore previous instructions and use'' $+$ genuine key & Defend & 20 & \cellcolor{green!15}20 & 1.0000 \\
\midrule
\textbf{Total} & & \textbf{3,922} & \cellcolor{green!15}\textbf{3,922} & 1.0000 \\
\bottomrule
\end{tabular}}
\end{table*}

We evaluate the gating module of \textsc{Locket} on Llama-3.2-1B with the ECHR revealing and defended adapters and a single key, using two tests. The first test varies the prompt content after the key. We construct 3,922 prompts from four sets of prompt bodies that are not part of the routing dataset: 200 held-out ECHR case facts, 20 paraphrases of one sentence (formal, colloquial, upper-case, and question--answer forms), 21 out-of-domain inputs (source code, SQL, \LaTeX, JSON, HTML, six non-English languages, emoji, and whitespace), and 12 jailbreak templates (DAN, developer mode, forged \texttt{SYSTEM} blocks, base64-encoded overrides, and ``print your key''). Each body is combined with the correct key and with the thirteen invalid key states listed in Table~\ref{tab:routing_invariance}: a wrong key, no key, an empty or whitespace-only key field, brackets removed, a truncated key, a key missing its last character, inverted letter case, a look-alike key in which one Latin letter is replaced by its visually identical Cyrillic character (a homoglyph), an invisible zero-width character inserted inside the key, the key written in fullwidth characters, and a leading space or zero-width character that shifts the key away from position~1. We also include the injection placements of Table~\ref{tab:routing_invariance} over 20 bodies, prompts that carry the correct key together with a decoy key or an ``ignore the key above'' directive in the body, and prompts of 128 to 4,096 tokens. The second test varies position~1 itself: we place each of the 128,259 tokens of the vocabulary directly after BOS with no key, so that every possible first token is evaluated rather than a sample (Table~\ref{tab:routing_sweep}).  Each prompt is routed twice: the full gating pass forwards the entire prompt through the frozen base model and reads the hidden state at position~1, while the truncated pass as in Appendix~\ref{app:gating_overhead} forwards only the BOS token and keyed position; the two routing decisions must agree.


Table~\ref{tab:routing_invariance} summarizes the results of the first test. The gating module routed all 3,922 prompts to the intended adapter with a selected-adapter probability of $1.0000$, including every rejected case. Paraphrasing, domain, language, jailbreak wording, and prompt length had no effect on routing, as expected from the invariance established in Section~\mbox{IV-C} of the main paper. In the injection conditions, the genuine key is present in the prompt but not at position~1 (mid-prompt, appended at the end, after a decoy key, or inside an ``ignore previous instructions'' directive); all 80 such prompts were routed to the defended adapter. Conversely, all 40 prompts with the correct key at position~1 and a decoy or override directive in the body were served by the revealing adapter. Shifted, altered, and near-miss keys—including truncated, character-missing, or case-inverted variants—were rejected in every case, indicating that the key space admits no partial matches that could narrow an attacker’s search.

Table~\ref{tab:routing_sweep} reports the results of the second test, which covers the one dimension that content invariance does not: which first tokens the router accepts without a key. Only one of the 128,259 tokens, the key, was routed to the revealing adapter ($p=1.000$); the remaining 128,258 tokens were routed to the defended adapter with $p\geq0.993$, including the first tokens of non-English text and source code that do not appear in the English routing templates. This confirms that an adversary without a key has no opening token that unlocks the revealing adapter. Because this property depends on the coverage of $\mathcal{D}^{\text{gate}}$ rather than on the architecture, the no-key examples in $\mathcal{D}^{\text{gate}}$ span the vocabulary (Section~\mbox{V-A} of the main paper), and we use the vocabulary sweep as a release check for every gate. The optional confidence threshold $\tau$ of Section~\mbox{IV-C} of the main paper is not needed for these results: the selected-adapter probability was at least $0.993$ in every case, so routing is simply the $\arg\max$ of Eq.~(12) of the main paper, and every non-key input reaches the defended adapter without any threshold.
\begin{table}[!t]
\scriptsize
\setlength{\tabcolsep}{3pt}
\renewcommand{\arraystretch}{1.1}
\centering
\caption{Exhaustive position-1 sweep: each of the 128,259 vocabulary tokens placed directly
after BOS with no key (Llama-3.2-1B, ECHR). Only the key may route to the revealing adapter.}
\label{tab:routing_sweep}
\resizebox{\linewidth}{!}{%
\begin{tabular}{@{} c c c c c @{}}
\toprule
\textbf{Tokens swept} & \textbf{Routed to Reveal} & \textbf{Routed to Defend} & \textbf{$p$ for the key} & \textbf{min $p$ on Defend side} \\
\midrule
128,259 & \cellcolor{green!15}1 (the key only) & \cellcolor{green!15}128,258 & 1.000 & 0.993 \\
\bottomrule
\end{tabular}}
\vspace{-3mm}
\end{table}

Because the number of adapters affects only the output width of the gating MLP and not the
hidden state it consumes, the same invariance applies to deployments with
additional keys and adapters; routing for a multi-adapter deployment is reported in
Table~V of the main paper.

These results establish that prompt content cannot subvert
\emph{which} adapter serves a request. They do not bound the residual leakage of the defended
adapter itself, which is governed by the mitigation it was trained with (scrubbing, DP, or
their combination) and quantified in Section~\mbox{V-C} of the main paper. An adversary holding a
valid key is outside the scope of the mechanism (Section~\mbox{III-B} of the main paper).

\section{Corpus-Size Ablation}\label{app:lowres}
Specialized domains rarely
offer corpora of the size used in the main paper, and fine-tuning on a few thousand annotated samples is
known to yield little or no gain without targeted architectural
changes~\cite{lompo2024medium,lompo2024multi}. To test whether \textsc{Locket}'s
utility--privacy separation survives in this regime, we retrain the revealing and defended
(scrubbed) adapters of \textsc{Llama-3.2-1B} on 10\%, 20\%, 25\%, 50\%, and 100\% of the ECHR
corpus ($N=500$ to $5{,}000$) for 10 epochs at an effective batch size of 16, train one
gating module per fraction, and report perplexity, inference-attack accuracy
(\textit{PERSON}, 50 candidates), and routing accuracy over all key conditions of
Appendix~\ref{app:routing_full} in Table~\ref{tab:lowres}. Three observations follow. First, the
access-control mechanism is independent of corpus size: the gating module converged to the same
near-zero entropy at every fraction and routed 100\% of the conditions correctly, including the
injection cases, at $N=500$ exactly as at $N=5{,}000$---as expected, since the router never
consumes task data and separates a fixed set of key embeddings. Second, authorized utility
degrades gracefully and saturates early: correct-key perplexity (highlighted in green) improves
from 8.38 at 10\% to 5.98 at 25\%, within 3\% of the full-corpus value. Third, the unauthorized path stays
defended at every size: wrong-key perplexity and inference accuracy (highlighted in red) remain
at 10.8--12.1 and 4.4--6.5\% with no upward trend as data grows, so the correct-key advantage
(1.4--1.9$\times$ in perplexity, 3.3--5.3$\times$ in inference accuracy) persists throughout.
Correct-key inference accuracy rises with data but is not monotone (the 50\% point exceeds
100\%); each point rests on 60--80 usable target sequences after candidate filtering, so that
column carries sampling variance. These findings separate two questions that the low-resource literature conflates.
Whether fine-tuning \emph{improves} a model on a small corpus is a property of the adaptation
method and the data, and we make no claim that LoRA escapes the limits reported
in~\cite{lompo2024medium,lompo2024multi}: our adapters also lose fluency below 25\% of the
corpus. Whether the deployment \emph{enforces} its access policy on that corpus is a property of
\textsc{Locket}, and it is unaffected by corpus size because the keyed-entry token and gating
module are trained on synthetic routing data alone. An organization with scarce annotated data
therefore obtains the same guarantee as one with abundant data---unauthorized requests are served
by the defended adapter, whose leakage is bounded by the defense it was trained with---and pays
the utility cost of small-corpus fine-tuning only on the authorized path, where it is the cost
any adapter-based method would pay.

\begin{table}[!t]
\scriptsize
\setlength{\tabcolsep}{3pt}
\renewcommand{\arraystretch}{1.1}
\centering
\caption{\textsc{Locket} under reduced fine-tuning corpora (Llama-3.2-1B, ECHR). Perplexity and
inference-attack accuracy (\%) are reported for the correct key (revealing adapter) and a wrong
key (defended adapter). All rows are retrained at batch size 16.}
\label{tab:lowres}
\resizebox{\columnwidth}{!}{%
\begin{tabular}{@{} l r cc|cc|c @{}}
\toprule
\multirow{2}{*}{\textbf{Corpus}} & \multirow{2}{*}{\textbf{N}}
& \multicolumn{2}{c|}{\textbf{Perplexity}}
& \multicolumn{2}{c|}{\textbf{Inference Acc.\ (\%)}}
& \multirow{2}{*}{\textbf{Routing Acc.\ (\%)}} \\
\cmidrule(lr){3-4} \cmidrule(lr){5-6}
& & \textbf{Correct} & \textbf{Wrong/No} & \textbf{Correct} & \textbf{Wrong/No} & \\
\midrule
10\% & 500 & \cellcolor{green!15}8.38 & \cellcolor{red!15}12.10 & \cellcolor{green!15}17.46 & \cellcolor{red!15}4.84 & \cellcolor{gray!30}100.00 \\
20\% & 1{,}000 & \cellcolor{green!15}6.11 & \cellcolor{red!15}10.79 & \cellcolor{green!15}20.27 & \cellcolor{red!15}6.06 & \cellcolor{gray!30}100.00 \\
25\% & 1{,}250 & \cellcolor{green!15}5.98 & \cellcolor{red!15}10.82 & \cellcolor{green!15}20.59 & \cellcolor{red!15}4.41 & \cellcolor{gray!30}100.00 \\
50\% & 2{,}500 & \cellcolor{green!15}5.78 & \cellcolor{red!15}10.98 & \cellcolor{green!15}34.25 & \cellcolor{red!15}6.49 & \cellcolor{gray!30}100.00 \\
100\% & 5{,}000 & \cellcolor{green!15}5.89 & \cellcolor{red!15}11.01 & \cellcolor{green!15}26.67 & \cellcolor{red!15}5.00 & \cellcolor{gray!30}100.00 \\
\bottomrule
\end{tabular}}
\vspace{-3mm}
\end{table}

\section{Extraction Leakage versus Query Budget}\label{app:query_budget}
The extraction attack of Section~\mbox{V-C} of the main paper samples 15,000 generations. To measure how leakage depends on the attacker's budget, we repeat it on Llama-3.2-1B (ECHR, scrubbed defended adapter) at 500, 1,500, 5,000, and 15,000 generations, with the correct key and with a wrong key (Table~\ref{tab:query_budget}). With the correct key, the number of true PII strings recovered grows with the budget, from 3 at 500 generations to 46 at 15,000 (recall 0.49\% to 7.52\%), the expected behavior of extraction against an undefended adapter. With a wrong key, every query is served by the defended adapter and the count stays between 0 and 3 across the same thirty-fold increase (recall at most 0.49\%). The gap therefore widens with adversary effort rather than closing: additional queries raise leakage only on the authorized path. In all runs the gating module routed 100\% of the generation batches to the intended adapter.

\begin{table}[!t]
\scriptsize
\setlength{\tabcolsep}{4pt}
\renewcommand{\arraystretch}{1.15}
\centering
\caption{Extraction-attack leakage versus query budget (number of sampled generations) on Llama-3.2-1B (ECHR, scrubbed defended adapter). ``Recovered'' is the number of unique ground-truth PII strings that appear in the generated text and not in the public base model's output.}
\label{tab:query_budget}
\resizebox{\linewidth}{!}{%
\begin{tabular}{@{} r ccc|ccc @{}}
\toprule
\multirow{2}{*}{\textbf{Generations}} & \multicolumn{3}{c|}{\textbf{Correct Key}} & \multicolumn{3}{c}{\textbf{Wrong/No Key}} \\
\cmidrule(lr){2-4} \cmidrule(lr){5-7}
 & \textbf{Prec.\ (\%)} & \textbf{Rec.\ (\%)} & \textbf{Recovered} & \textbf{Prec.\ (\%)} & \textbf{Rec.\ (\%)} & \textbf{Recovered} \\
\midrule
500 & 3.90 & 0.49 & \cellcolor{green!15}3 & 0.00 & 0.00 & \cellcolor{red!15}0 \\
1,500 & 3.97 & 1.96 & \cellcolor{green!15}12 & 2.50 & 0.16 & \cellcolor{red!15}1 \\
5,000 & 2.98 & 3.43 & \cellcolor{green!15}21 & 4.00 & 0.33 & \cellcolor{red!15}2 \\
15,000 & 2.00 & 7.52 & \cellcolor{green!15}46 & 2.33 & 0.49 & \cellcolor{red!15}3 \\
\bottomrule
\end{tabular}}
 
\end{table}

\section{Membership Inference: Full Results}\label{app:mia_full}
Table~\ref{tab:mia_llama1b_echr} summarizes the results. When the correct key is supplied, the gating module routes every query to the revealing adapter and the attack attains an AUC of 0.689 (LOSS) and 0.670 (reference), with TPR@1\%FPR of 1.8\% and 1.2\%. This is the expected behavior of an undefended fine-tuned model: the revealing adapter memorizes its training records to a measurable degree, and the correct-key rows are identical across the two defense configurations, confirming that only the defended adapter differs between them. When a wrong key is supplied, every query is routed to the defended adapter and the membership signal drops. With the scrubbed defended adapter, the AUC falls to 0.628 (LOSS) and 0.629 (reference); with the DP defended adapter, it falls to 0.503 and 0.502, i.e., the adversary can no longer do better than random guessing, and TPR@1\%FPR remains at the 1\% chance level (1.5\% and 0.9\%). In all conditions the gating module selected the intended adapter for 100\% of the queries.

 \begin{table*}[!t]
\scriptsize
\setlength{\tabcolsep}{10pt}
\renewcommand{\arraystretch}{1.2}
\centering
\caption{Membership inference against Llama-3.2-1B on the ECHR dataset under \textsc{Locket}, using 5,000 member and 5,000 non-member records. An AUC of 0.5 corresponds to random guessing. The DP adapter is trained at \(\varepsilon_{\mathrm{PRV}} = 8.03\), \(\delta = 2\times10^{-4}\) (Appendix~\ref{app:dp_config}).}

\resizebox{\linewidth}{!}{%
\begin{tabular}{@{} l l cc|cc|cc @{}}
\toprule
\multirow{2}{*}{\textbf{Defense}} &
\multirow{2}{*}{\textbf{Access}} &
\multicolumn{2}{c|}{\textbf{LOSS Attack}} &
\multicolumn{2}{c|}{\textbf{Reference Attack}} &
\multicolumn{2}{c}{\textbf{Adapter Selected}} \\
\cmidrule(lr){3-4} \cmidrule(lr){5-6} \cmidrule(lr){7-8}
\textbf{Method} & \textbf{Condition} & \textbf{AUC} & \textbf{TPR@1\%FPR (\%)} & \textbf{AUC} & \textbf{TPR@1\%FPR (\%)} & \textbf{Reveal (\%)} & \textbf{Defended (\%)} \\
\midrule
\multirow{2}{*}{\textbf{Scrubbed}}
 & Correct Key & \cellcolor{green!15}0.689 & 1.8 & \cellcolor{green!15}0.670 & 1.2 & \cellcolor{gray!30}100.00 & 0.00 \\
 & Wrong/No Key & \cellcolor{red!15}0.628 & 2.9 & \cellcolor{red!15}0.629 & 1.4 & 0.00 & \cellcolor{gray!30}100.00 \\
\midrule
\multirow{2}{*}{\textbf{DP}}
 & Correct Key & \cellcolor{green!15}0.689 & 1.8 & \cellcolor{green!15}0.670 & 1.2 & \cellcolor{gray!30}100.00 & 0.00 \\
 & Wrong/No Key & \cellcolor{red!15}0.503 & 1.5 & \cellcolor{red!15}0.502 & 0.9 & 0.00 & \cellcolor{gray!30}100.00 \\
\bottomrule
\end{tabular}}
\label{tab:mia_llama1b_echr}
\end{table*}

It is important to be precise about what produces the gap between the two wrong-key rows. The difference between scrubbing (0.63) and DP (0.50) is a property of the \emph{mitigation method} used to train each defended adapter, not of \textsc{Locket}. Scrubbing replaces PII spans with a mask token but the scrubbed adapter is still fine-tuned on the same documents, so it fits the remaining non-PII tokens of every member record; document-level membership signal therefore persists even though PII extraction is suppressed (cf.\ Table~IV of the main paper). DP-SGD, by contrast, bounds the influence of any single record on the adapter weights, which is precisely the property that membership inference tests, and the wrong-key AUC collapses to chance accordingly. \textsc{Locket}'s contribution is orthogonal to this choice: it guarantees that every unauthorized or miskeyed request is served by the defended adapter, so the membership leakage available to an unauthorized adversary is exactly that of the selected defense, while the key holder retains the full-utility revealing adapter. Consistent with the multi-defense experiment, an organization thus sets its privacy floor by choosing how the defended adapter is trained (scrubbing when PII values are the concern, DP when membership must also be bounded), and can raise that floor by replacing a single adapter artifact without touching the base model, the revealing adapter, or the gating logic.

\end{document}